\documentclass[article,11pt]{emulateapj}
\usepackage{graphicx}
\usepackage{times}
\usepackage{natbib}
\usepackage{amsmath}
\usepackage{amssymb}
\def\apjs{Astrophys. J. Supp.}
\def\grl{Geophys. Rev. Let.}

\def\planss{Planet. \& Space Sci.}

\def\asht{AsH$_3$ }

\def\hto{H$_2$O }

\def\deg{$^\circ$ }
\def\degx{$^\circ$}

\def\mum{$\mu$m }
\def\mumx{$\mu$m}

\def\pht{PH$_3$ }

\def\pthf{P$_2$H$_4$ }
\def\pthfx{P$_2$H$_4$}
\def\nht{NH$_3$ }

\def\nhfsh{NH$_4$SH }
\def\nhfshx{NH$_4$SH}

\begin{document}

\title{Interpretation of shadows and antishadows on Saturn and the evidence
against south polar eyewalls.}

\author{L.A. Sromovsky$^1$, P.M. Fry$^1$, and K.H. Baines$^1$}
\affil{$^1$Space Science and Engineering Center, University of Wisconsin-Madison,
1225 West Dayton Street, Madison, WI 53706, USA}

\slugcomment{Journal reference: L.A. Sromovsky, P.M. Fry and K.H. Baines, 
Icarus, https://doi.org/10.1016/j.icarus.2019.113399}
\begin{abstract}
Cassini spacecraft observations of Saturn in 2006 revealed south polar
cloud shadows, the common interpretation of which was initiated by
Dyudina et al. (2008, Science 319, 1801) who suggested they were being
cast by concentric cloud walls, analogous to the physically and
optically thick eyewalls of a hurricane.  Here we use radiative
transfer results of Sromovsky et al. (2019, Icarus, doi.org/10.1016/j.icarus.2019.113398), in
conjunction with Monte Carlo calculations and physical models, to show
that this interpretation is almost certainly wrong because (1)
optically thick eyewalls should produce very bright features in the
poleward direction that are not seen, while the moderately brighter
features that are seen appear in the opposite direction, (2) eyewall
shadows should be very dark, but the observed shadows create only
5-10\% I/F variations, (3) radiation transfer modeling of clouds in
this region have detected no optically thick wall clouds and no
significant variation in pressures of the model cloud layers, and (4)
there is an alternative explanation that is much more consistent with
observations.  The most plausible scenario is that the shadows near
87.9\deg S and 88.9\deg S are both cast by overlying translucent aerosol layers
from edges created by step decreases in their optical
depths, the first in the stratospheric layer
at the 50 mbar level and the second in a putative diphosphine layer
near 350 mbar, with optical depths reduced at the poleward side of
each step by 0.15 and 0.12 respectively at 752 nm.  These steps are
sufficient to create shadows of roughly the correct size and shape,
falling mainly on the underlying ammonia ice layer near 900 mbar, and
to create the bright features we call antishadows. 
\end{abstract}
\maketitle
\shortauthors{Sromovsky et al.}
\shorttitle{South polar shadows and antishadows on Saturn.}
\pagebreak

\section{Introduction}

Both polar regions of Saturn are characterized by strong cyclonic
vortices, inferred by tracking cloud features
\citep{Sanchez-Lavega2006,Baines2009cyclone,Antunano2015} and by applying the thermal
wind equation to temperature structure retrieved from Cassini thermal
infrared observations \citep{Fletcher2008}.  \cite{Fletcher2008} also
found a small depletion of \pht within a few degrees of both poles,
suggesting local downwelling motions.  The generally low optical
depths inferred for cloud layers in the polar region
\citep{Baines2018GeoRL,Sro2019spole} also suggest a general polar
downwelling, as do the changes they found in the \pht profile and decline in the \asht mixing
ratio towards the pole.  The polar regions also have some of the morphological
characteristics associated with hurricanes in Earth's atmosphere,
including dark eyes, which themselves indicate reduced particulate
scattering that would be expected from downwelling motions inside the
eye regions.  In the case of the south polar region, there is also
evidence of cloud shadows.  \cite{Dyudina2008Sci,Dyudina2009}
suggested that the shadows were cast by a double eyewall cloud
structure. They also inferred a deep vertical convection extending
over two scale heights, based on the detection of the shadow-producing
cloud boundaries in methane band images.

However, if these Saturnian eyewalls are regions of dense clouds
produced by deep convection, such convection might be expected to
produce lightning, as is present in earthly eyewalls and in two other
instances of deep convection on Saturn, namely in the Great Storm of
2010-2011 \citep{Dyudina2013} and in the thunderstorms of Storm Alley
\citep{Dyudina2007}.  In both of those cases, there was also spectral
evidence of unusual convective activity provided by Cassini's Visual
and Infrared Mapping Spectrometer (VIMS).  That evidence was the
appearance of optically thick clouds that also displayed the 3-\mum
absorption signature of \nht ice \citep{Baines2009stormclouds,Sro2013gws,Sro2018dark}.
These signatures would not have been detected without the convection
of ammonia ice particles from deeper levels to the top of the overlying main
upper tropospheric cloud, which otherwise over most of Saturn shields
the ammonia ice layer from detection by external observations.
Although a 3-\mum absorption signature is seen in the south polar
region \citep{Sro2019spole}, it is not evidence for deep
convection. Instead, \cite{Sro2019spole} explained its appearance by
the dramatically lower optical depth of the upper tropospheric cloud
in the polar region.  They also inferred cloud structure at multiple
latitudes within the region containing the two eyewall features, but
failed to find any clouds of high optical density, or any obvious
dramatic changes in cloud structure in the vicinity of the observed
shadows. They also showed that optical depths of cloud layers
decreased toward the pole within the eye region, and that the decrease
took the form of small step changes in the top tropospheric layer.  Our
usual methods of radiative transfer are invalid in close proximity to
such changes because they rely on horizontally homogeneous
conditions. Thus, we here use Monte Carlo calculations to investigate
the cloud boundaries.

In the following, we review the observations of cloud shadows,
and add an additional observational constraint, which
is provided by observations of bright features produced
near the same cloud boundary that produces shadows. 
We should expect bright features to be produced by eyewalls
when they are illuminated by sunlight at low angles, but
the observed bright features extend in the wrong direction
and are more consistent with what we will define as
antishadows, which are produced by thin translucent particle
layers overlying a deeper scattering layer.  After a qualitative
review of the observations, and a more detailed comparison with
radiative transfer modeling outside the shadows, we then describe our 
approach for Monte Carlo modeling of the I/F profile at and
near boundaries that produce both shadows and antishadows, which
is the simplest method that can treat such boundary problems.
We then describe the results of applying Monte Carlo calculations
to interpret eyewall and thin layer boundary profiles. That
is followed by a summary of our conclusions.

\section{Cassini ISS observations of south polar clouds in 2006}

\subsection{Shadows and antishadows}

\begin{figure*}[!hbt]\centering
\includegraphics[width=6.2in]{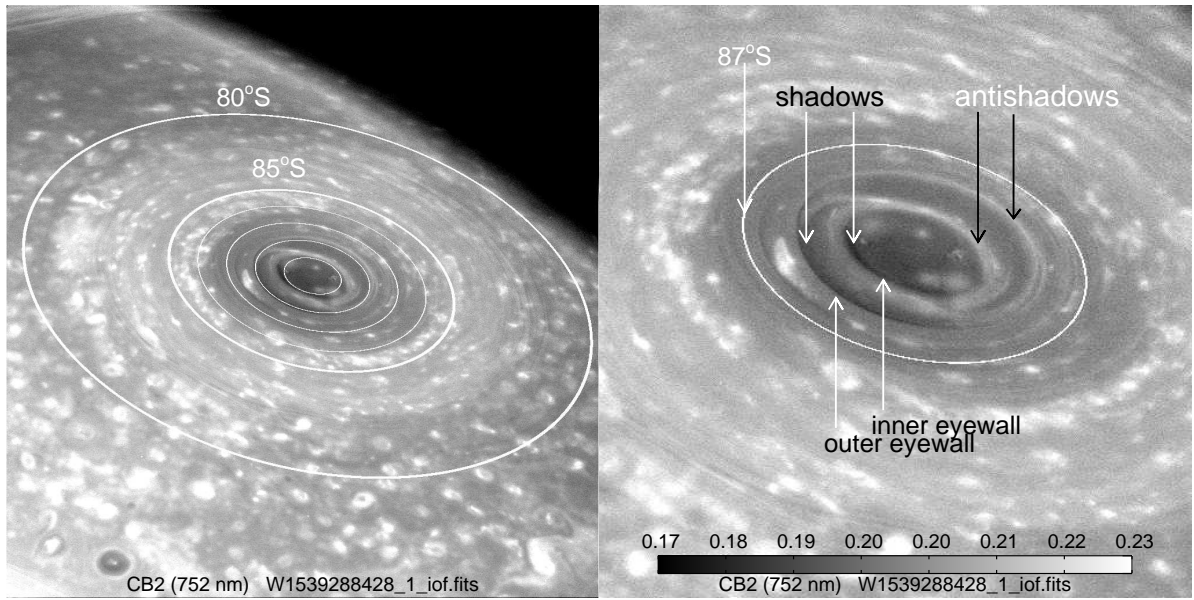}
\caption{Normal (left) and enlarged (right) views of Saturn's south pole as captured in an ISS
image taken with a CB2 (752 nm) filter on 11 October 2006. Planetocentric latitude lines are 
shown in the left panel for 80\degx S, and at 1\degx intervals from 85\degx S to 89\degx S.  Sunlight was incident
from the lower left when this image was taken. Cloud boundaries near 89\degx S and 88\degx S
are the features commonly described as inner and outer eyewalls respectively.  The features we define to be
antishadows are indicated in the right panel.}\label{Fig:polecontext}
\end{figure*}

We first turn our attention to selected Cassini/ISS observations of
the south polar region of Saturn acquired during October of 2006, which
are summarized in Table \ref{Tbl:issobs}. The images were processed and
navigated as described by \cite{Sro2019spole}.  The morphology of this region
is illustrated by the ISS 752-nm image shown in Fig.\ \ref{Fig:polecontext},
in which I/F values are corrected for limb darkening using a Minnaert function of the form,
\begin{eqnarray}
  I/F [x,y]_C  = I/F [\mu(x,y),\mu_0(x,y)]\times \nonumber \\
(\mu_{0P}(x,y)/\mu_{0})^{K-1}(\mu_P(x,y)/\mu)^K,
\end{eqnarray}  
where $I/F(x,y)_C$ is the I/F value at image coordinates (x,y)
corrected to the observing geometry at the south pole, $\mu(x,y)$ and
$\mu_0(x,y)$ are the observer and solar zenith angle cosines at the
same image point, subscript $P$ indicates values at the image center
(near the pole), and $K$ is the Minnaert exponent, which we adjusted
to a value of 0.72.  At the wavelength of this image, the dark eye
region extends from about 86\deg to the pole.  Outside the eye, the
polar latitudes are peppered with small bright cloud features,
suggestive of fair weather cumulus.  These features are almost
completely absent from the eye region.  There is a further darkening
in the inner eye, which extends roughly from 89\degx S to the
pole. Note the dark crescent-shaped shadows, which mark what is
generally referred to as the outer and inner eyewalls.  As will be
made more obvious in polar projections, both of the eyewall boundaries
are oval rather than circular, and they rotate about the pole at
different rates.  Also note that on the opposite side of the pole from
the shadows are bright features that we labeled as antishadows.

\begin{table*}[!htb]\centering
\caption{Cassini ISS images we used and their observing conditions.}
\setlength\tabcolsep{3pt}
\begin{tabular}{ l c c c c c r}
                             &                   &     UT Date  &        Start &   Pixel &    Phase & Fig. \\
ISS image ID               &  ISS Filter      & {\footnotesize YYYY-MM-DD}  & Time & size &    angle      &    Ref.  \\
\hline
       W1539288428  &    CB2 + CL2 (752 nm)           &  2006-10-11 & 19:35:09    & 17.5 km &  37.5\deg & \ref{Fig:polecontext}, \ref{Fig:shadow}, 3, \ref{Fig:mcmodcomp} \\
       W1539290817  &    MT2 + CL2 (728 nm)   &  2006-10-11 & 20:14:56    & 17.1 km & 32.6 \deg &  \ref{Fig:mcmodcomp}\\
       W1539293298  &    MT3 + CL2 (890 nm)   &  2006-10-11 & 20:56:08    & 33.6 km & 27.4\deg & \ref{Fig:polarmeth} \\
       W1539293315  &    CB2 + CL2 (752 nm)  &  2006-10-11 & 20:56:37  & 16.8 km & 27.3\deg & \ref{Fig:shadow},
                 \ref{Fig:polarmeth}\\
       W1539293355  &    MT2 + CL2 (728 nm)  &  2006-10-11 & 20:57:14    & 16.8 km & 27.3\deg & \ref{Fig:polarmeth}\\
       W1539298642  &    CB2 + CL2 (752 nm)  &  2006-10-11 & 22:52:23  & 16.4 km & 16.9\deg & \ref{Fig:shadow}, 3\\
\hline
\end{tabular}\label{Tbl:issobs}
\end{table*}

\begin{figure*}[!htb]\centering
\includegraphics[width=6.5in]{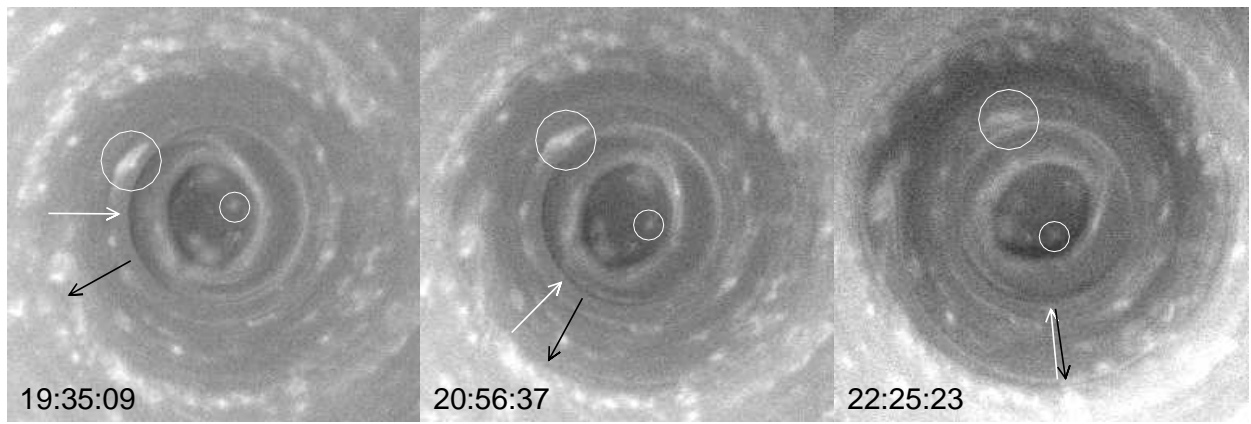}
\caption{Polar projections of a sequence of ISS 752-nm images of
  Saturn's south polar region on 11 October 2006. Note that shadows follow the
  direction of incident sunlight, indicated by the white arrow in each
  panel. The black arrows point toward the Cassini spacecraft. The
  latitude circle tangent to the boundaries of each image has a radius
  of 5\deg planetocentric. The zero of longitude is at the right.}\label{Fig:shadow}
\end{figure*}

The evidence that the dark crescents in the eye region are in fact shadows comes 
from their orientation relative to the direction of incoming sunlight.  As the
planet rotates, the shadows move in step with the longitude of the sun, as
illustrated in Fig.\ \ref{Fig:shadow}, where incoming sunlight is shown by white
arrows and black arrows point towards the Cassini spacecraft.  Note also the
bright antishadow associated with the outer ``eyewall'' moves with the sun
as well, staying on the opposite side of the pole from the sun, and replacing
the poleward shadow seen at the left in the first panel with an anti-poleward
antishadow near the top of the third panel.  The bright cloud feature just above
the antishadow in the third panel (encircled by a large circle in each
subpanel of Fig.\ \ref{Fig:shadow}) has only moved only about 30\deg of longitude between
the first and third panels, while the encircled smaller bright feature near the inner ``eyewall''
has moved about 80\deg over the same interval.  That feature also has moved
more in longitude than the other features inside the inner eye and appears
to be near the peak inner vortex angular motion profile measured by \cite{Dyudina2009} and
\cite{Antunano2015}.  Also note the elongated oval shape of the inner eyewall. Both of these inner eyewall
features are also moving faster than bright features outside the eye region.

\begin{figure}[!htb]\centering
\includegraphics[width=3.5in]{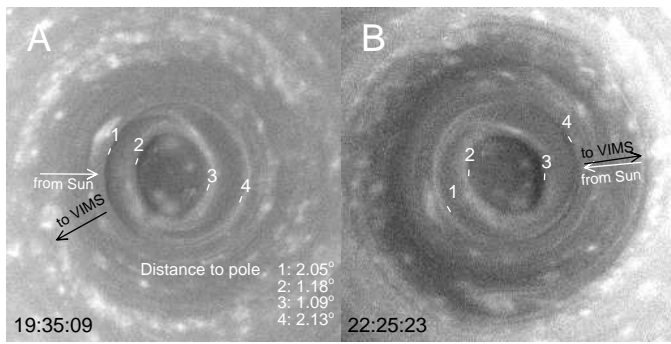}
\caption{Close-up view of first (A) and last (B) polar projections
  from Fig.\ \ref{Fig:shadow}. The putative eyewalls are marked by
  short white arcs labeled 1-4. These are located at longitudes
  200\degx E (1 and 2) and 20\degx E (3 and 4) in A, but in panel B
  are shifted 40\degx E (1 and 4) and 73\degx E (2 and 3) due to zonal
  motions.  The B image is shown rotated counterclockwise 90\deg from
  its position in Fig.\ \ref{Fig:shadow} to put the cloud features in
  roughly the same orientation as in A, at a time when the sun is
  incident from roughly the opposite direction. Note in particular the
  effects of sun reversal at boundary 3, which in A displays a relatively
  bright antishadow to its right with sunlight incident from the left,
  but in B, the antishadow disappears and the same edge casts a shadow
  to the left resulting from sunlight here coming from the right.  If
  the shadow had been cast by a wall, that wall should appear to VIMS
  as a bright feature in A, extending to the left, opposite to what is
  observed.}\label{Fig:closeup}
\end{figure}

\begin{figure}[!t]\centering
\includegraphics[width=3.5in]{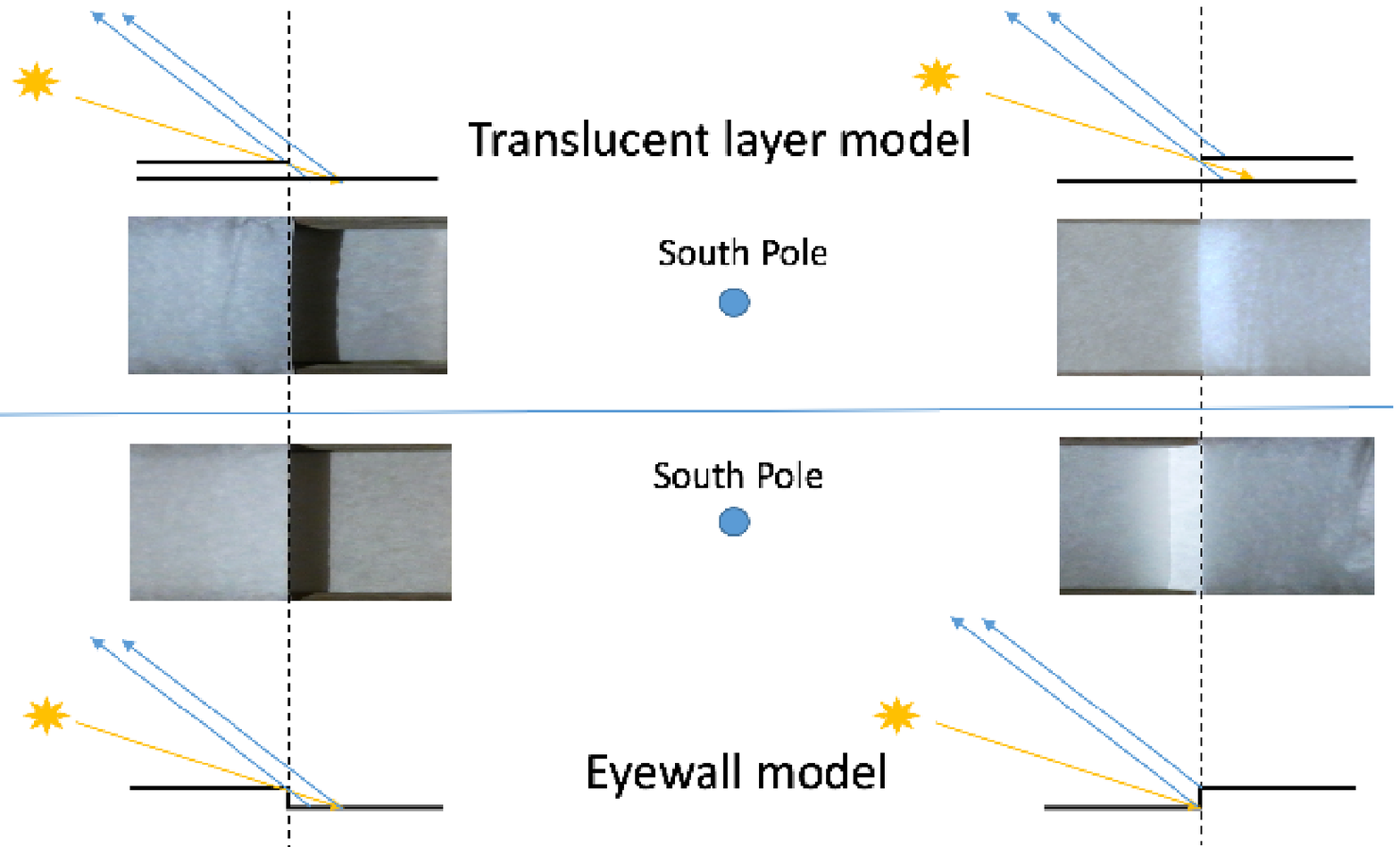}
\caption{Photographs of tracing-paper physical models illustrating two
  alternative shadow production mechanisms: a sharp change in optical
  depth of a translucent layer (top) or a step change in cloud
  pressure at an eyewall edge (bottom). Light from the sun is
  indicated by yellow rays, and light to the observer is indicated by
  blue rays.  Each model produces a shadow on the lower layer when the
  transition is located on the sun-ward side (left) of the pole.  But
  on the opposite side (right) the translucent layer model displays a
  moderately bright ``antishadow'' produced by light shining
  underneath the top layer and providing extra illumination from
below, while the alternate model displays a
  brightly illuminated eyewall.  When illuminated and observed at the angles
  illustrated here, which are comparable to those for the ISS images
  in Fig.\ \ref{Fig:shadow}, the antishadow is seen to extend from
  the upper layer boundary (marked by dashed lines) away from the
  pole, which is also the direction seen in the ISS images, while the
  bright eyewall is seen to extend towards the pole, in conflict with
  ISS observations. Figure from \cite{Sro2019spole}.}
\label{Fig:physmod}
\end{figure}

The behavior of shadows and antishadows is easier to see in an enlarged
direct comparison between earliest and latest observations, as shown
in Fig. \ref{Fig:closeup}.  In this case the third panel of
Fig.\ \ref{Fig:shadow} is rotated 90\deg counterclockwise to put the
incident sun direction in that panel almost exactly opposite
to that shown in the first image, while roughly retaining the
same orientation of cloud features.  Here, the solar elevation is near
25\degx, while the observer (VIMS) elevation angle is near 56\degx, which
makes it possible to see whatever shadows are created, if dark enough and
long enough.  Shadow boundaries, marked by short arcs, appear in both
images at the same cloud feature locations in both images (they track
the zonal flow).  When these boundaries
are between the pole and the sun, they cast shadows toward the pole,
but when they are on the opposite side of the pole from the sun,
they cast antishadows away from the pole.  If the shadows had been
cast by vertical wall clouds, then in the opposite sun direction
 the walls would have been visible as bright
features extending towards the pole.

The distinction between these two alternate shadow-producing mechanisms
is perhaps more easily appreciated with photographs of a toy
physical model presented in Fig.\ \ref{Fig:physmod}.  At the top 
are photographs of a translucent separated layer model, constructed
of tracing paper and shown in cross section above or below
the corresponding photographs. The model was photographed in two different positions,
corresponding to locations on opposite sides of a model pole. The incident sunlight is indicated by
yellow rays and the blue rays point to the observer (VIMS). The sun
and observer angles are comparable to those of the VIMS observations
in Fig.\ \ref{Fig:closeup}. On the left
a shadow is produced by the top layer of tracing paper, beginning where
its optical depth drops to zero.  On the right, an antishadow is produced
by sunlight shining underneath the edge of the upper layer resulting
in extra illumination from below, which is visible from above because
the layer is not opaque.

For the eyewall model in Fig.\ \ref{Fig:physmod}, a shadow is produced
when the wall is on the left, but on the right, the illuminated eyewall
is visible to the observer and appears to extend towards the pole from
the top boundary of the wall.  The wall is much brighter than the top
or bottom surfaces because they are illuminated at a low elevation,
while the wall is illuminated at nearly normal incidence.  This effect
is also seen for hurricane eyewalls on Earth under similar observing conditions
as we later illustrate in Sec.\ \ref{Sec:eyewall}, along with
Monte Carlo calculations showing similar results.

\begin{figure*}[!htb]\centering
\includegraphics[width=5in]{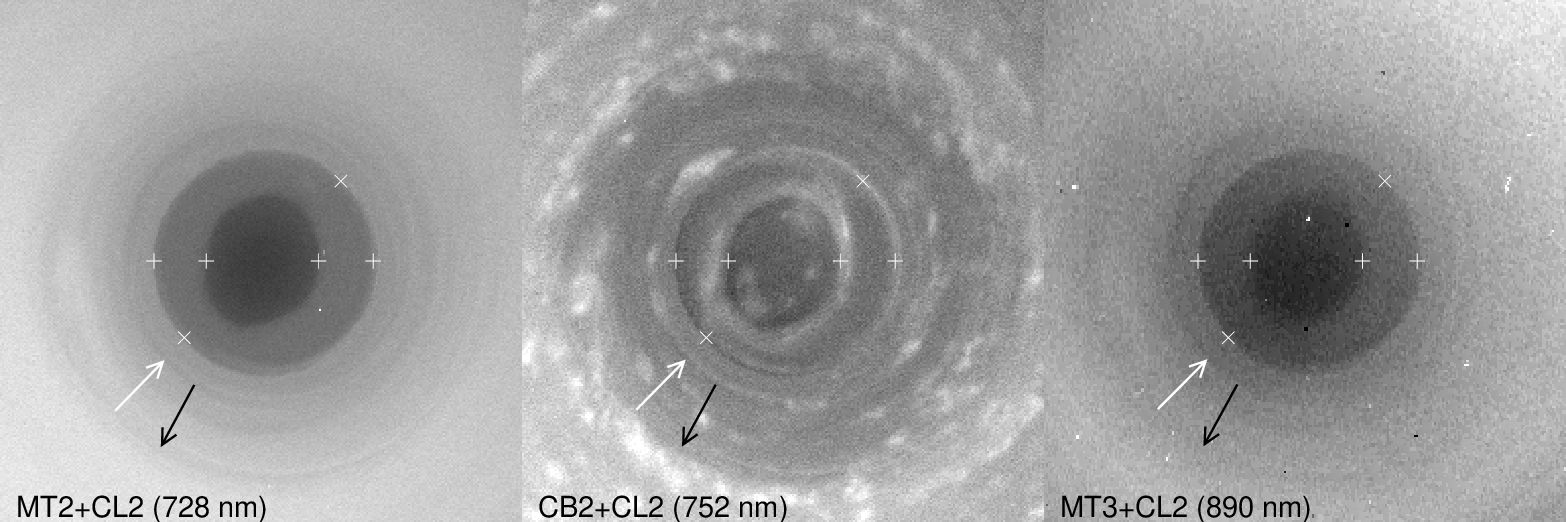}
\caption{Comparison of methane band images (MT2 at left and MT3 at
  right) and continuum ISS image (center) of Saturn's south polar
  region on 11 October 2006, presented as a polar projection extending
  $\pm$5\deg from the pole.  The central ISS image was taken at
  20:56:37.0 UT using a CB2 (752 nm) filter, and has an image scale of
  16.8 km/pixel at the center of the image. Its identification number
  is W1539293315. The other images were taken 37 seconds earlier (890
  nm image) and 37 seconds later (728 nm).  The direction of incoming
  sunlight is indicated by white arrows and the direction of light
  traveling to the Cassini spacecraft is indicated by black
  arrows. Fiducial marks at the same set of locations in each image
  show that boundaries in the methane band images correlate with
  shadow and antishadow features in the continuum images.  These cloud
  boundaries have come to be thought of as inner and outer eyewalls.}\label{Fig:polarmeth}
\end{figure*}

\subsection{The evidence for deep convection}

\begin{figure*}\centering
\includegraphics[width=5.2in]{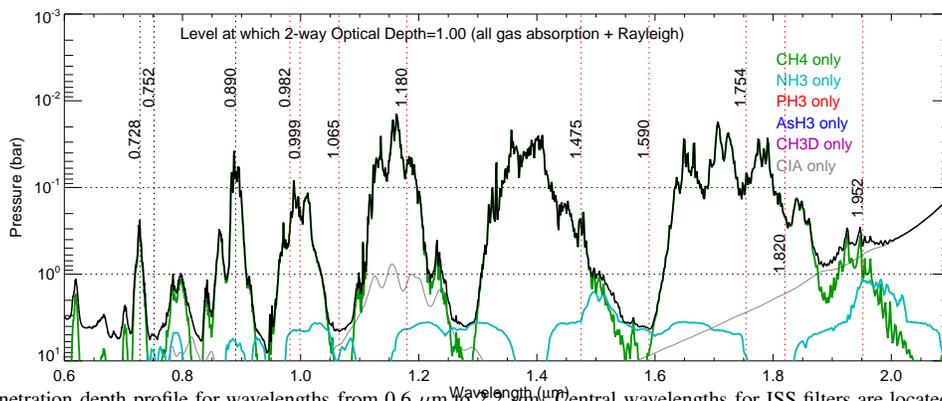}
\vspace{-0.2in}
\caption{Vertical penetration depth profile for wavelengths from 0.6 \mum to
  2.2 \mumx. Central wavelengths for ISS filters are located by
  vertical dotted lines. VIMS wavelengths spanning a similar range of
  penetration depths are indicated by vertical red dotted lines.}\label{Fig:pdblowup}
\end{figure*}

The evidence cited by \cite{Dyudina2009} for the deep convection over
two scale heights, reaching higher than the tropopause, is the
visibility of the eyewall clouds in methane band images. Although they
did not display and annotate the relevant images in their paper, it
seems clear that it is the boundaries displayed in
Fig.\ \ref{Fig:polarmeth} that form the basis for their inference.
This figure displays polar projections of ISS images taken with
methane band filters MT2 (728 nm) and MT3 (890 nm), in comparison with
a polar projection of an image taken with the CB2 (752 nm) filter.
The depth of penetration of these images can be roughly assessed from
Fig.\ \ref{Fig:pdblowup}.  At vertical viewing the two-way optical
depth in the MT2 and MT3 filters are reached at pressure of 250 mbar
and about 60 mbar respectively, the latter reaching above the tropopause
($\sim$70 mbar) even without accounting for the low view angle for the observations in
question, which would lower these pressures by about a factor 3.1 ($\mu_0$ = 0.25, $\mu
= 0.5$). The CB2 filter on the other hand, reaches 2-way absorption optical
depth unity at about 2.3 bars for the same observing and illumination geometry.  

Fiducial
marks on each image in Fig.\ \ref{Fig:polarmeth} are placed at the same locations in latitude and
longitude.  The methane band images both show what appears to be a
stair-step decrease in I/F towards the pole, with step changes
occurring at the same locations as the boundaries associated with
cloud shadows seen the the CB2 image in the middle panel.  Neither
methane band image actually shows either of the bright rings that
might be associated with the top of an isolated eyewall. But in a
hurricane seen from above, there is a thick cirrus outflow that blocks our
view of the wall clouds.  If that is happening in the polar region of
Saturn, one would expect to see optically thick clouds extending
somewhat away from the pole starting at each of the two boundaries.
One would expect the composition of that ``cirrus shield'' to be the
same as that of the deep convective cloud, which, if the Great Storm
analysis is taken as a guide, might contain a mix of ammonia ice,
water, and perhaps \nhfsh \citep{Sro2013gws}, or even \pthfx. But there
is no evident 3-\mum absorption that would indicate the presence of either
\nht or \nhfshx.  On the other hand, it is
conceivable that the boundaries seen in the methane image are actually
due to the changing character of upper cloud layers, which is not
connected to deep convection at all, {\it but is instead the cause of the
cloud shadows from which the deep convection has been inferred}.

\subsection{Possible radiative transfer constraints}

There are two sorts of constraints that can
 further guide our interpretation of these features.  One is the analysis of
near-IR spectra in the south polar region to constrain the vertical cloud
structure.  Such an analysis might reveal the presence of optically thick clouds
extending to high altitudes, or define the locations of layers on which
shadows might be cast, or define changes in layer scattering properties
with latitude, to see if the sudden disappearance of one layer at key
latitudes could produce a shadow on a lower layer.  There were in fact
near-IR observations by the Visual and Infrared Mapping Spectrometer (VIMS)
taken during the same period covered by the ISS images, and an analysis by
\cite{Sro2019spole} provides useful constraints on the models. While useful,
these constraints are unable to be applied within shadows or close to
shadow or antishadow boundaries, because they rely on the existence of
horizontally homogeneous conditions. Thus they need to be supplemented by
a radiative transfer analysis that can model conditions at and near such
boundaries.  For that we use Monte Carlo calculations.  But these are limited
in a different way.  Because of the time involved in computing ray paths
and interactions for many millions of photons, it is not practical to
include the complexities of gas absorption models.  Thus Monte Carlo
calculations are mainly useful only for continuum spectral regions and pressure
ranges, where gas absorption can be largely ignored. First, we consider the
results of the VIMS analysis.

\section{Vertical cloud structure inferred from Cassini VIMS}

\subsection{Overview of Cassini VIMS Near-IR results}

 The VIMS instrument \citep{Brown2004} combines
two mapping spectrometers, one covering the 0.35-1.0 \mum spectral range,
and the second covering an overlapping near-IR range of 0.85-5.12
\mumx. The near-IR spectrometer use 256 contiguous channels sampling
the spectrum at intervals of approximately 0.016 \mumx.  The near-IR VIMS
spectra are particularly useful in constraining the composition of Saturn's
cloud layer because many of the candidate cloud materials have near-infrared 
absorption features, especially in the vicinity of 3 \mumx.
From several studies of VIMS spectra in different regions on Saturn,
including the Great Storm of 2010-2011 \citep{Sro2013gws} and its wake
region \citep{Sro2016}, the Storm Alley
region \citep{Baines2009stormclouds,Sro2018dark}, the north polar region \citep{Baines2018GeoRL},
and the south polar region \citep{Sro2019spole}, the following picture
of Saturn's multi-layer cloud structure has emerged.

Below a generally thin stratospheric haze, Saturn is covered by a
thick cloud layer of unknown composition. This material seems to have
no near-IR absorption features, although it may have features that are
located within, and masked by, gas absorption features due to
phosphine and ammonia.  The prime candidate for this layer of
particles is diphosphine.  But we cannot test this hypothesis with
spectral observations because the optical properties of diphosphine
have not yet been quantitatively characterized.  Another candidate
that is far less likely on photochemical grounds is some form of white
phosphorus.  Lacking a dipole moment, a lack of significant near-IR
absorption features would be expected for phosphorus.

Below this tropospheric mystery layer is a more well established layer of
condensed ammonia.  This material can be observed spectrally, but over
most of Saturn its 3-\mum spectral features
 can only be seen in areas of intense convection,
such as in Storm Alley, or in the Great Storm of 2010-2011.  Another
place where the spectral features of \nht ice can be seen is in polar
regions where the main upper tropospheric mystery cloud has
dramatically reduced optical depth.

The deepest layer that can be sensed using visible to near-IR
wavelengths is likely composed of \nhfshx.  Where convection is
vigorous, e.g., where lightning is observed, water ice may also
be a significant cloud component. The deep clouds need to
provide significant optical depth at pressures less than 5 bars in order
to modulate the level of 5-micron emission observed on Saturn. If
assumed to be optically dense, cloud top pressures in the 2-4 bar
range are often found.   Ammonia clouds
could also modulate this emission to some degree, but in most cases the ammonia
layer, as constrained by scattered sunlight observations, is too
optically thin to provide much blocking of thermal emissions
in the 4.8-5.1 \mum region.  

Radio observations have shown that lightning on Saturn has been highly
variable over time, but highly localized in space.  Lightning has only
been observed in the Great Storm of 2010-2011, and in Storm Alley. No
lightning has been observed in the south polar region, which makes it
unlikely that there is any vigorous convection, even within the cloud
structures identified as eyewalls.

\subsection{Near-IR results for south polar vertical cloud structure}

\begin{figure*}[!hbt]\centering
\includegraphics[width=5in]{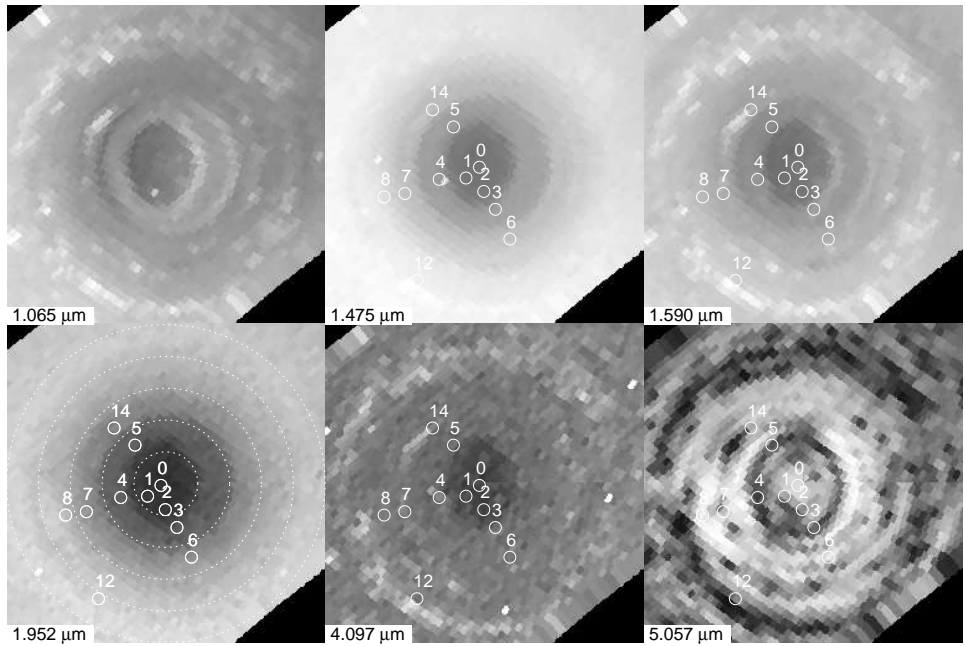}
\caption{VIMS images of the south pole of Saturn at six wavelengths
  indicated in the panel legends. These images (from cube
  V1539288419\_1), were taken at 19:35:01 UT on 11 October 2006 at a
  phase angle of 37.2\deg and with a sub-spacecraft pixel size of 145
  km. Processing and calibration details are described by
  \cite{Sro2019spole}.  The penetration depths of the wavelengths less
  than 2.2 \mum are indicated by 2-way optical depth profiles in
  Fig.\ \ref{Fig:pdblowup}.  The clear-atmosphere penetration depths
  at 4.1 \mum and 5.06 \mum are given by \cite{Sro2019spole} and are
  about 1 and 3 bars respectively, which is where the two-way optical depth at
  normal viewing reaches unity.  Numbered circles indicate the locations where
  complete near-IR spectra were analyzed by \cite{Sro2019spole}, with
  results plotted in Fig.\ \ref{Fig:vimsfits}.  These polar
  projections provide complete coverage out to latitude 85\degx S,
  with latitude circles at 1\deg intervals in the 1.952-\mum image.}\label{Fig:irpole}
\end{figure*}

The Cassini Visual and Infrared Mapping Spectrometer (VIMS) also observed
Saturn's south polar region in 2006. Polar projections at selected wavelengths
are displayed in Fig.\ \ref{Fig:irpole}. At the continuum wavelength of 1.065 \mumx,
which is similar in penetration depth and appearance to images taken with the
ISS CB2 (752 nm) filter but at an order of magnitude worse spatial resolution,
we see clear shadow and antishadow features, with somewhat muted versions
seen in the continuum wavelength of 1.59 \mumx. Lowered contrast for both
features at longer wavelengths would be expected if they are produced by an optically thin overlying
layer of small particles, because optical depths would decrease with wavelength.
Eyewall clouds, on the other hand, would not exhibit much wavelength dependence
because of their significant optical depths.  

\begin{figure}[!hbt]
\hspace{-0.15in}\includegraphics[width=3.5in]{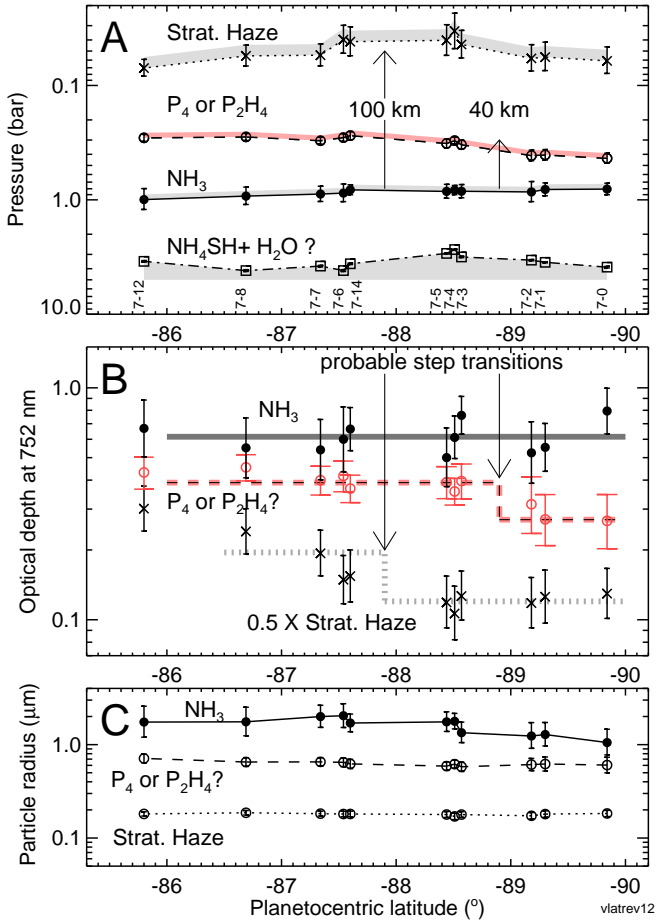}
\caption{Best-fit parameter values vs. latitude for cloud structure
  fits to background cloud spectra between 85.8\degx S and the south
  pole, including pressures (A), optical depths at 752 nm (B), and
  particle size (C).  The solid gray horizontal bar in (B) is at
  optical depth 0.615, the dashed step is from 0.39 to 0.27 optical
  depths at 1.1\degx from the pole, and the dotted step is
  from 0.39 to 0.24 optical depths at 2.1\degx from the
  pole, but plotted at half scale to eliminate overlap with the upper
  layer.  From \cite{Sro2019spole}.}\label{Fig:vimsfits}
\end{figure}

\cite{Sro2019spole} derived model 
cloud structures at locations marked by numbered circles in Fig.\ \ref{Fig:irpole} from the same VIMS 
observations (but using complete near-IR spectra, not just sample wavelengths).  Note that
these samples avoid small discrete cloud features and mostly stay away from shadow and antishadow
boundaries.  The results from their radiative transfer modeling of these background
cloud spectra are plotted in Fig.\ \ref{Fig:vimsfits}.
Unlike the dramatic reduction in optical depths found by \cite{Baines2018GeoRL}
 within the eye of the north polar vortex, the changes seen as the south pole is
approached are relatively modest.  All four layers are preserved, but the stratospheric
and putative diphosphine layers declined towards the poles,  as might be expected in the case of
downwelling motion.   It is noteworthy that the \nht layer optical depth remained
relatively constant for the background cloud structure, and none of the layers became completely
cleared out within the eye region. 

The top of the deep putative \nhfsh + \hto layer varies between 3 and
4 bars, exhibiting more spatial structure than evident in the
overlying layers.  From 86.2\degx S to 89.8\degx S, the ammonia layer
pressure smoothly decreases from about 1 bar to 800 mbar, and its
optical depth (at 752 nm) appears to be constant over that range, at a
value of 0.615, but with a substantial uncertainty of about 25\% in
individual fit results.  The putative \pthf layer moves slightly
downward as the pole is approached, from about 300 mbar to 430 mbar,
while its optical depth declines from about 0.39 to 0.27, likely
occurring mostly in a step decrease of 0.12 at 88.9\degx N, the location of
which is based on ISS images.  The effective pressure of the
stratospheric haze decreases then increases as the pole is approached,
with a log slope vs latitude comparable to that seen in the putative
\pthf layer.  Its optical depth decreases toward the pole, most likely
occurring mainly in a step change from 0.39 to 0.24 at 87.9\degx S.
The particle radii for the stratosphere and putative \pthf layers
average about 0.18 \mum and 0.6 \mum respectively, with very little
change versus latitude.  The ammonia layer particle radius declines
from about 1.9 \mum to 1.1 \mum as the pole is approached.

The upper two layers dip downward in altitude near the pole, while the
ammonia layer rises upward slightly, suggesting that the polar
downwelling is limited to the upper two layers.  It also appears that
the upper two layers are plausible sources of shadows on the ammonia
layer, and perhaps to some degree on the deeper layer.  Whether the
optical depth declines in the stratospheric haze and in the putative
diphosphine layer (upper tropospheric layer) are smooth functions of
latitude or in the form of sudden step changes could not by determined
from the radiative transfer modeling of VIMS spectra, both because the
model assumption of horizontally homogeneous conditions do not apply
at the transitions, and, even if they did, the uncertainty in
retrieved parameter values is large enough to obscure the transition
details.  However, \cite{Sro2019spole} did conclude that the
transitions were in the form of step changes at in the stratospheric haze and \pthf
layers because images sensing those layers predominantly (mainly using
the MT2 filter) did show step changes in I/F, as is evident in our
Fig.\ \ref{Fig:polarmeth}, and in the VIMS images at 1.475 \mum and
1.952 \mumx, which are shown in Fig.\ \ref{Fig:irpole}.  While the
VIMS spectral modeling results did not discover anything that might be
considered evidence for deep convective eyewalls, they did provide
evidence for declining optical depth and, with the help of imaging
observations, evidence that the decline is almost certainly in steps.

In the following, we will show that step changes in optical depth of a
translucent overlying layer can produce shadows on a lower layer and
antishadows visible at the top of the shadow casting layer. We will
also show that if eyewalls were present they would indeed create
bright features that would replace shadows as the sun's direction
reversed, but the location of those bright features is in conflict
with observations.  We begin with a description of our Monte Carlo
calculation methods.

\section{Monte Carlo radiative transfer methodology}

\subsection{Motivation and limitations}

Our doubling and adding routines for radiative transfer assume that
atmospheric gas and aerosol structures are horizontally homogeneous.
Practically speaking, this means that the horizontal variation scale
is much larger than the vertical scale.  Since the vertical scale of
the cloud structure is of the order of 100 km, and the horizontal
variation with the eye region is of the same scale in many locations,
and even smaller scales at shadow boundaries, we cannot use our normal methods to do
accurate radiative transfer modeling in the vicinity of shadows.  To
model I/F profiles in the vicinity of cloud boundaries, we thus turned
to Monte Carlo calculations, which can handle such situations.
However, the time consuming nature of such calculations requires us to
simplify the problem by working only at wavelengths for which
atmospheric absorption can be neglected.  At 752 nm, atmospheric
absorption is minimal, reaching a vertical 2-way absorption optical
depth of unity at about the 8-bar level.  But at the typical slant
angles of the subject observations, the zenith angle cosines are
$\mu_0$ = 0.25 for illumination and $\mu$ = 0.55 for viewing, so that
the absorption optical depth of unity is reached at p = 8 bar
/(0.5/$\mu_0$ + 0.5/$\mu$), which evaluates to 2.3 bars.  At the top
of the \nht layer, which is near 800 mb, the two way absorption
optical depth would only be about 0.35.  The importance of this
absorption would be amplified by reflections between layers, but this
effect is not as important for optically thin layers that lose photons
before many reflections can occur.  To better assess the importance of
ignoring atmospheric absorption in our Monte Carlo calculations, we
ran a full radiative transfer model calculation for a background cloud
feature at 88.6\degx S using parameters in column 5-3 of Table 4 of
\cite{Sro2019spole} and using the same radiative transfer code, then
computed a spectrum for the same aerosol model but with the methane
mixing ratio reduced by a factor of 10.  This only produced a 10\%
increase in the I/F at 752 nm for the appropriate observing and
illumination conditions.  Thus, we are confident that our neglect of
atmospheric absorption at 752 nm will not seriously impact our Monte
Carlo calculations.

\subsection{Vertical and horizontal structure model}

\begin{figure}[!htb]
\hspace{-0.05in}\includegraphics[width=1.43in]{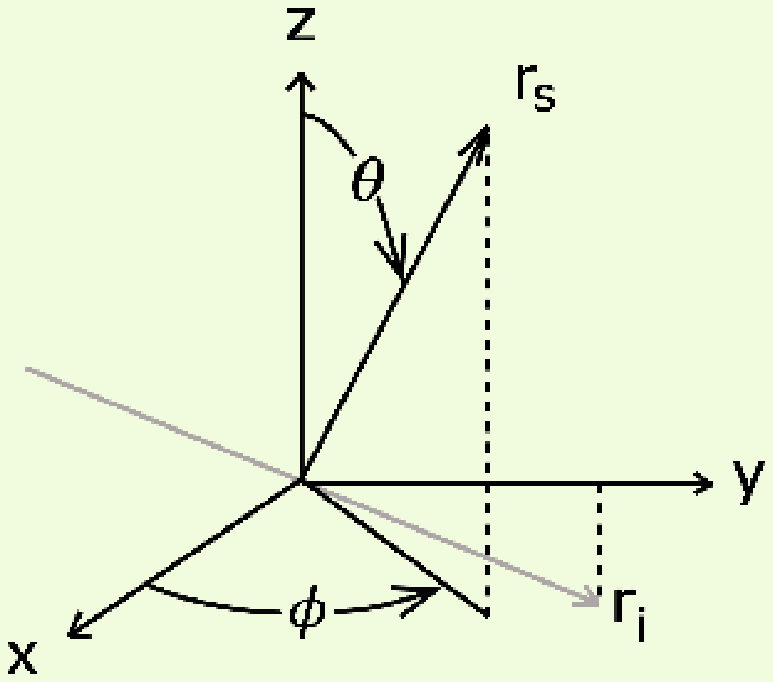}
\includegraphics[width=1.98in]{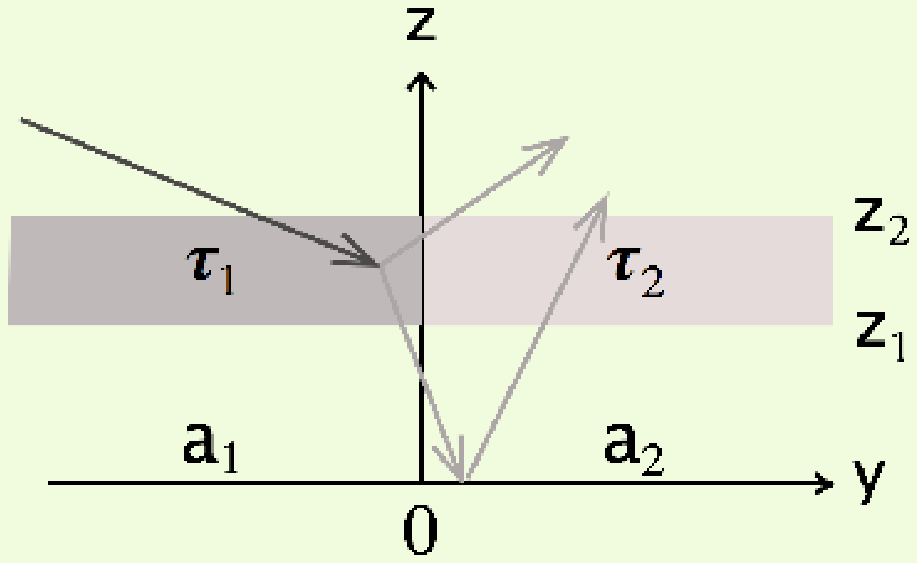}
\caption{Coordinate system for Monte Carlo calculations (left), where
vector $r_i$ indicates a typical incident ray in the $y-z$ plane,
and vertical structure model (right), shown in cross section,
with a sample incident ray (in black) and a few possible scattered rays (in gray)
following a random interaction sequence.  For these examples the only scattering
occurs in the layer between $z_1$ and $z_2$ or at the surface, both of
which are uniform in the $x$ direction but can have different
scattering properties for $y<0$ and $y\ge 0$. The $x$-$y$ positions of
incident and exit rays are measured where they intersect the
$z = z_2$ plane.}\label{Fig:coord}
\end{figure}

Our basic aerosol structure model consists of a layer of
finite (and adjustable) thickness defined by upper and lower
boundaries $z_1$ and $z_2$, suspended at a height $z_1$ over a Lambertian
scattering surface at $z = 0$.  The scattering layer is three
dimensional, but its properties are invariant in the $x$ dimension.
Our coordinate system and the simplified model are illustrated
in Fig.\ \ref{Fig:coord}.  All properties are assumed invariant in
the $x$ direction, but a discontinuity in suspended layer properties and
surface reflectivity are allowed at $y$ = 0.  For the suspended layer
we assume that optical density, i.e., optical depth per unit length,
is independent of altitude between $z_1$ and $z_2$, defined by the vertical
optical depth $\tau$ divided by the layer thickness $(z_2-z_1)$.  For
$y <$ 0 we assume $\tau = \tau_1$, single scattering albedo $\varpi = \varpi_1$, and a surface albedo $a_1$,
while for $y >$ 0 we assume $\tau = \tau_2$, $\varpi = \varpi_2$,
 and surface albedo $a_2$.  This allows for absorption by the
aerosol particles and by the surface. 
The scattering and absorption by the atmosphere, as noted earlier, are ignored. 

\subsection{The Monte Carlo sampling equations}

A key element of Monte Carlo radiative transfer is randomly sampling
scattering parameters, some of which can have non-uniform probability
distributions $P(x)$, such as optical depth and scattering angle. To do this
with a random number generator that produces values uniformly
distributed between 0 and 1, we make use of the cumulative probability
distribution function, given by
\begin{eqnarray}
 P_C(x) = \int_{xmin}^xP(x')dx'/\int_{xmin}^{xmax}P(x')dx',
\end{eqnarray}
where $P_C(x)$ varies from 0 to 1 as $x$ ranges from $xmin$ to $xmax$.
If $P_C$ is set equal to uniformly distributed random
values between 0 and 1, and then inverted to find
the corresponding values of $x$, it is easy to see that
 these $x$ values will be distributed
in proportion to the slope $d P_c(x)/dx$ = $P(x)$, which provides
the desired probability distribution of $x$.
In some cases $P_C(x)$ can be computed analytically, while in
others, it is necessary to compute numerical integrals to
create a table of $x$ and $P_C(x)$ values.  In either case a random
value of $x$ is obtained by solving for $x$ in the equation
\begin{eqnarray}
 \xi = P_C(x)
\end{eqnarray}
where $\xi$ is selected randomly from a distribution that
is uniform between 0 and 1.  To sample optical depths
between zero and infinity, we use
\begin{eqnarray}
\tau = -\log_e(1- \xi) \label{Eq:tau}
\end{eqnarray}
To sample an azimuthal angle and zenith angle cosine from an isotropic
scattering phase function, we use
\begin{eqnarray}
\phi_s = 2\pi\xi_1\\
\mu_s = 2\xi_2 - 1 
\end{eqnarray}
where the different subscripts on $\xi$ indicate that two unrelated random
samples are chosen for the two directional parameters.  For
scattering according to a Henyey-Greenstein (HG) function, there is
an analytical function for sampling the cosine of the scattering angle \citep{Witt1977},
which is given by
\begin{eqnarray}
\mu_s = \frac{1+g^2-[(1-g^2).(1-g+2g\xi )]^2}{2g} \quad \mathrm{for~} g \ne 0\\
\mu_s = 1-2\xi  \quad \mathrm{for~} g = 0
\end{eqnarray}
However, for double H-G functions or Mie scattering we are reduced to
interpolation of numerical tables to find the value of $\mu_s$ for a
given sample of $\xi$.  To facilitate tracing of the scattered ray,
the direction specified by $\mu_s$, which is the cosine of the
scattering angle, and $\phi_s$, which is an azimuth angle of the
scattered ray measured in a plane perpendicular to the incident
vector, must be expressed in terms of cosine $\mu$ and angle $\phi$ as defined
in the original coordinate system.
 
\subsection{The Monte Carlo ray tracing algorithm}

Our basic approach is along the lines described by
\cite{Whitney2011}.  We choose a specific direction for incoming sunlight, specified by
a zenith angle cosine $\mu_0$, and an azimuth angle $\phi_0$, which is
the angle of the vector projection in the $x$-$y$ plane measured
from the $x$ axis counter clockwise about the $z$ axis (Fig.\ \ref{Fig:coord}).  
Because the incoming light from the sun is a downward vector in this coordinate
system its zenith angle cosine $\mu_0$ is negative.
In what follows, we only consider cases in which the incoming
light is in the $y$-$z$ plane.  Next, a random
incident location uniformly distributed between $y=$-L and $y=$+L is chosen.
Recall that y positions are measured at the top of the scattering layer (the
$z = z_2$ plane).  (Although the scattering
layer is three dimensional and photons are scattered in three dimensions,
because there is no variation of layer properties in the $x$ direction,
we choose all incident rays to be incident at $x=0$ and ignore
$x$ variation in the output results.)  Our next step is to choose
a random optical depth using Eq.\ \ref{Eq:tau}. This
optical depth is then converted to a physical distance over which the photon
travels before interacting with a scattering particle or the surface.
Inside the scattering layer we find a distance by
 dividing by the optical density (optical depth per unit length).
If the length times $\mu_o$ exceeds the layer thickness, the photon
will exit the bottom of the layer and hit the surface. Otherwise it
will scatter inside the layer.

If the photon scatters inside the layer, then it either gets absorbed
within the layer with a probability 1-$\varpi$, or it scatters in a random
direction.  If it is absorbed it will not be counted.  If it scatters,
the direction will be randomly chosen in accord with the assumed phase
function, using the above sampling equations.  If the particle passes
through the upper layer and reaches the surface, it may be absorbed or
reflected.  If the surface albedo is less than 1, its probability of
being reflected is equal to either $a_1$ or $a_2$, depending on
whether it impacts the surface at $y < 0$ or $y > 0$.  What happens to
a specific photon is determined by taking a uniformly distributed
sample $\xi$. If $\xi$ is less than 1-$a_1$, for example, then the
photon is absorbed and that photon is not counted.  Otherwise the
photon is reflected, and then a random reflection direction is chosen,
using $\phi = 2\pi \xi_1$ and $\mu = \sqrt{\xi_2}$. The latter relation
can be obtained from the constraint that the radiance of a
Lambertian reflector is independent of observing angle.

Once a new location and direction are determined for the photon, a new
optical depth is sampled, the photon path is traced to its next
interaction, following the same procedures.  The ray tracing is
complicated by the fact that there are two different regions with
different properties, i.e. for $y<0$ and $y \ge 0$ (see Fig.\ \ref{Fig:coord}).  
If the photon is not absorbed, it will eventually exit the top of the scattering
layer, at which point its exit location and direction will be saved
for later binning and computation of radiance and I/F.  The uncertainty of a
binned value is taken to be the square root of the number of photons
in the bin.  For reasonable accuracy without excessive computer time
we generally used between 20 and 100 million incident photons. Computation
time for a given case ranged from 20 minutes to an hour or more.

\subsection{Computing radiance and I/F from binned photon counts}

For a given observer viewing geometry specified by $\phi_v$ and $\mu_v$,
and chosen respective bin sizes $d\phi_v$ and $d\mu_v$, we find array
locations for all the photons falling inside the viewing bins and then
compute a histogram of all those photons as a function of spatial
location $y$ at which photons are incident at the top of the
scattering layer, computing their counts in bins of size $dy$.  If
$N_{phot}$ is the total number of photons that are incident, then the
photon flux is the number of photons per unit area orthogonal to the
incident direction, i.e.  $F_\mathrm{phot} =
N_\mathrm{phot}/(|\mu_0|(\Delta y\Delta x))$, where $N_\mathrm{phot}$ is
the total number of photons that were incident (per unit time is implicit),
 $\Delta y=2L$, and $\Delta x$ is some arbitrary
distance in the $x$ direction (all variation in the $x$ direction is
ignored).  The photon radiance for a Lambertian surface so illuminated
is then given by
$R_\mathrm{Lambert} = F_\mathrm{phot}/\pi$.  The observed photon
radiance is given by $R_\mathrm{photon} = dN_\mathrm{phot}(y,\mu_v,\phi_v)/(dy \Delta x d\mu_v d\phi_v
\mu_v)$, where $dN_\mathrm{phot}(y,\mu_v,\phi_v)$ is the number of photons in bins
$y\pm dy/2$, $\mu_v\pm d\mu_v/2$, and $\phi_v\pm
d\phi_v/2$.  The reflectivity is then computed as I/F = photon
  radiance/Lambert radiance, in which the arbitrary interval $\Delta x$
cancels out:\begin{eqnarray}
I/F =  \pi\: |\mu_0|\: \times \nonumber \\dN_\mathrm{phot}(y,\mu_v,\phi_v)/
[ N_\mathrm{phot} \mu_v d\mu_v d\phi_v dy/\Delta y].
\end{eqnarray}

\subsection{Code validation}

The code used for the Monte Carlo calculations was tested first by
setting model parameters on both sides of the boundary to the same
values and then comparing those results with doubling and adding
computations for the same vertical structure and particle scattering
parameters.  We next used different parameter sets for the two regions
of the Monte Carlo code and verified that away from the joint the
Monte Carlo results approached the values obtained for horizontally
uniform conditions.  Although we could not directly verify radiance
calculations near the discontinuity in parameter values (i.e. at and near
$y=0$), we did verify that the ray tracing was being done correctly,
by following rays as they crossed into different regions to verify
that correct branches were chosen in the computation of scattering
probabilities and directions.  Finally, as a sanity check, we built a
physical model that generated shadows and antishadows, and were able
to produce similar effects with the Monte Carlo code.

\section{Monte Carlo calculation results}

\subsection{Scattering properties of eyewalls}\label{Sec:eyewall}

\begin{figure*}[!htb]\centering
\begin{minipage}[b]{2.51in}
\includegraphics[width=2.5in]{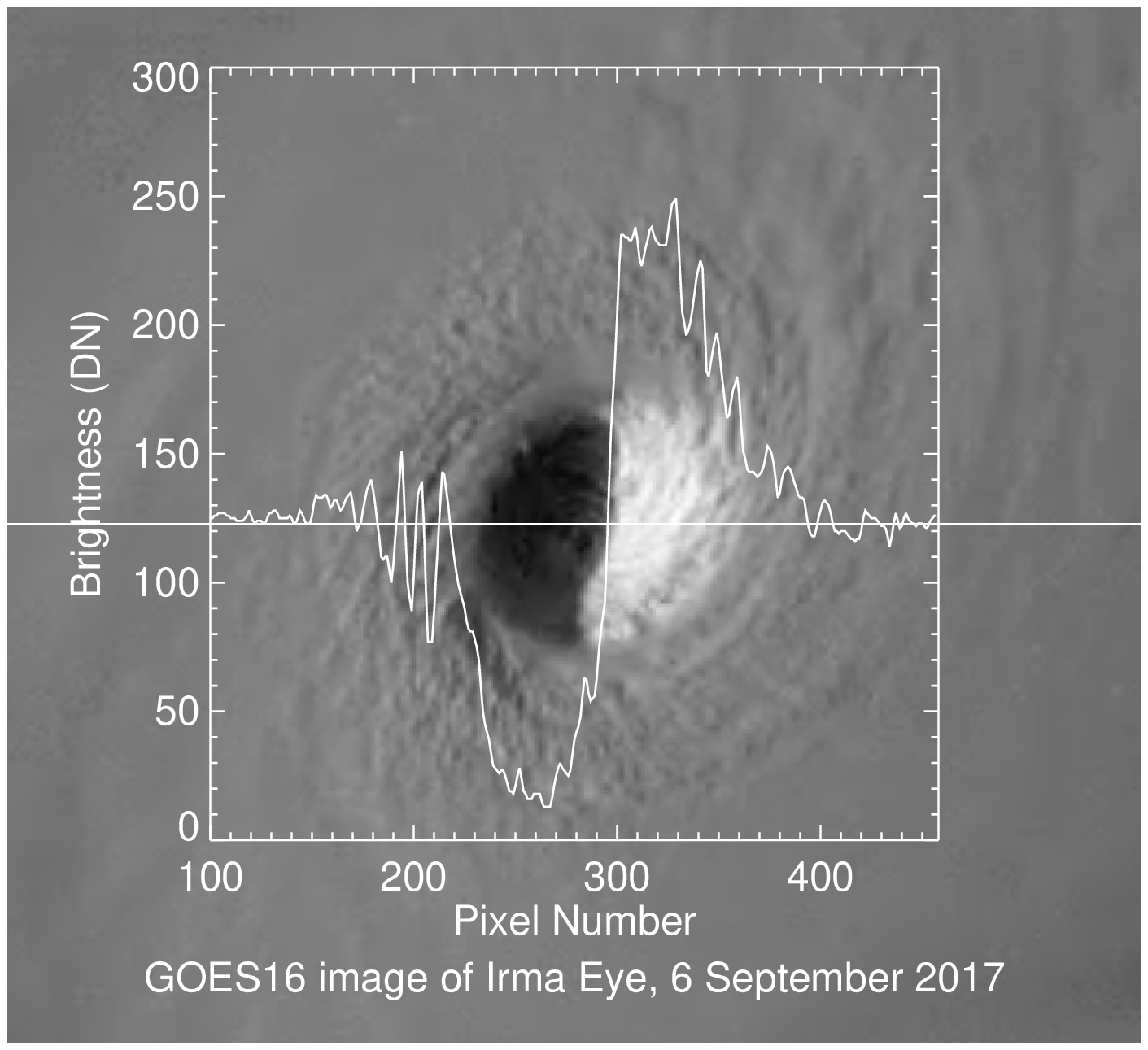}
\includegraphics[width=2.5in]{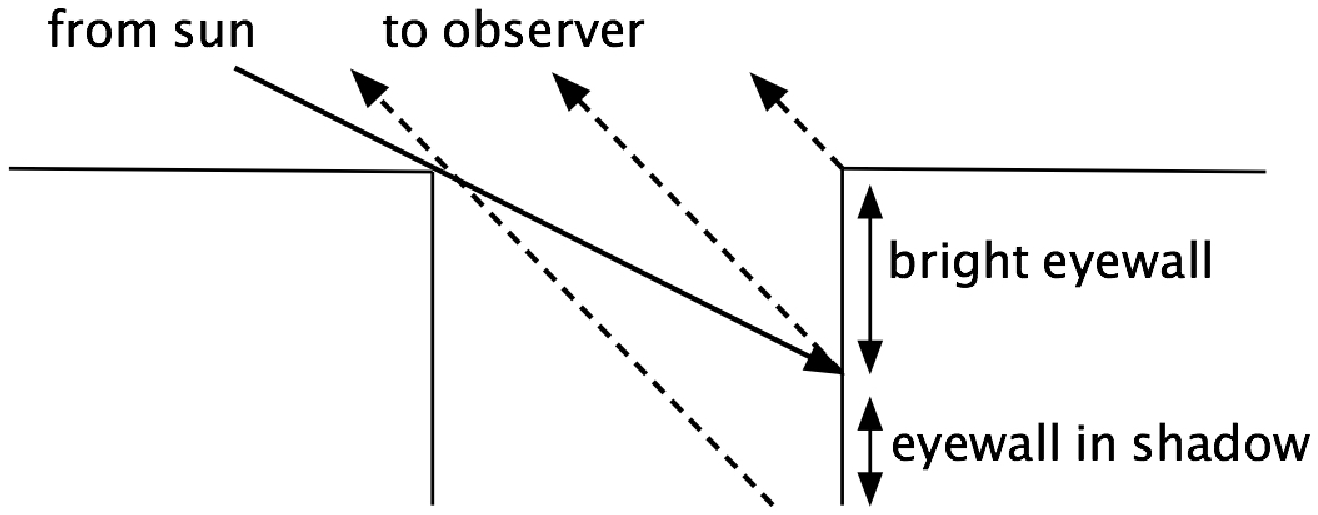}
\end{minipage}\hspace{-0.08in}
\includegraphics[width=3.75in]{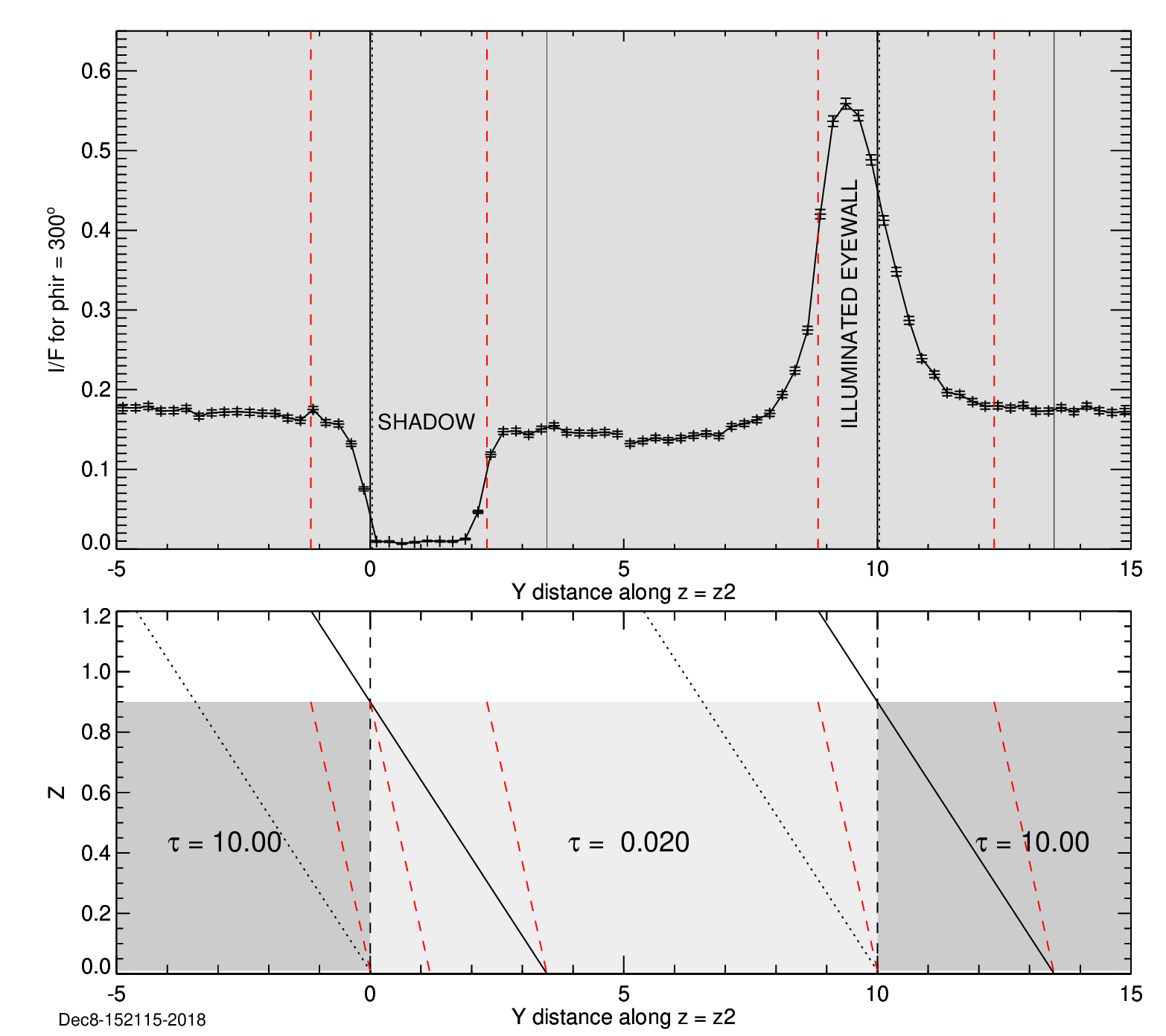}
\caption{{\bf Left:} The eye of hurricane Irma at low sun angles
  displays an illuminated eyewall that is much brighter than the top
  cloud layer of the hurricane. The illumination and observing
  geometry is shown below the image and a scan across the circular eye
  displays the brightness profile. {\bf Right:}. A similar brightness
  profile is seen in Monte Carlo calculations for Saturn's south polar
  region under 2006 Cassini observing conditions. This calculation is
  for an eye with a width much greater than its height, so that the
  shadow does not fall on the eyewall.  The diagram in the lower part
  of the right panel displays incoming solar rays with black lines and
  observer sight lines with dashed red lines. The wall clouds have a
  vertical thickness of 0.9 units of distance and a separation of 10
  units. The illumination is at $\mu_0 = -0.25$ and the sight line is
  at $\mu_v$ = 0.55, with $\phi_0 = 90$\deg and $\phi_v =
  270$\degx. The observer is thus able to see the eyewall shadow on
  the left and the illuminated eyewall on the right.  The I/F profile
  across this structure is shown in the top panel, computed for
  conservative cloud particles of radius 5-\mumx, refractive index of 1.82+0$i$,
  and vertical optical depth of 10.  At the bottom of the model we
  assumed a Lambertian surface of albedo = 0.6}\label{Fig:irma}
 \end{figure*}

The term eyewall we take to imply optically thick vertical
convective towers similar to what is observed in hurricanes on Earth.
If not heavily obscured by cirrus outflow clouds, eyewalls
would be able to cast shadows on either side of the pole, both
when the sun was incident toward the pole and when it was incident
away from the pole. As shadows
on Saturn are only observed to extend toward the pole,  either the eyewalls
are obscured by cirrus outflow, or they do not exist.
Another property of earthly eyewalls is that
when illuminated at low sun angles, they are very bright in comparison to
the top visible cloud layer, which becomes dark as the sun
sets.  If we approximate both the top and wall clouds as
Lambertian reflectors of unit albedo, the brightness
of the top would be proportional to $\mu_0 = \cos{\theta_0}$, while
that of the wall would be proportional to $\sin{\theta_0}$, so the
ratio of the wall I/F to the top I/F is roughly proportional
to the tangent of the solar zenith angle, yielding a ratio
of 3.9 to 1 at $\theta_0$ = 75.5\degx.
 An example of this effect can be seen in the left
panel of Fig.\ \ref{Fig:irma},
which displays the eye of hurricane Irma at a low sun angle.
A scan through the center of the eye, shows that the
shadow is very deep, only about 10\% of the brightness of
clouds outside the wall, while the illuminated eyewall
is about double the brightness of background clouds.  

A Monte
Carlo calculation for a typical south polar Saturn viewing
geometry is shown in the right panel of Fig.\ \ref{Fig:irma}.  Note that
the I/F at the top of the wall, away from the eye region is about
0.18, while in the shadow it is near zero and in the illuminated
eyewall it peaks near 0.55, which is about 3.1 times
the background I/F (a little below the ratio of 3.9 given by
the Lambertian approximation described
in the previous paragraph). The peak - background difference
of 0.37 is 2.2 times the background - shadow difference of 0.17.
Redoing the calculation with a particle radius of 0.5 \mum instead
of 5.0 \mumx, yielded a slightly larger peak and slightly larger
background, but a similarly large difference ratio, in this case
2.45 instead of 2.2. Particle radii between
0.5 and 5 \mum gave intermediate results.  Reducing the optical
depth of the eyewall clouds by a factor of five reduced the
peak I/F to 0.35 (for a 1-\mum particle), which is still almost twice the background
I/F of 0.18, although the differential difference ratio dropped to
about 1.0.  Yet, the location of the peak is still inside the
eye, rather than on the top of the wall cloud.  It should be noted that
an optical depth of 2.0 is too low to be considered characteristic of
a deep convective eyewall. 

\subsection{Scattering properties of translucent layers}

\begin{figure}[!htb]\centering
\includegraphics[width=3.5in]{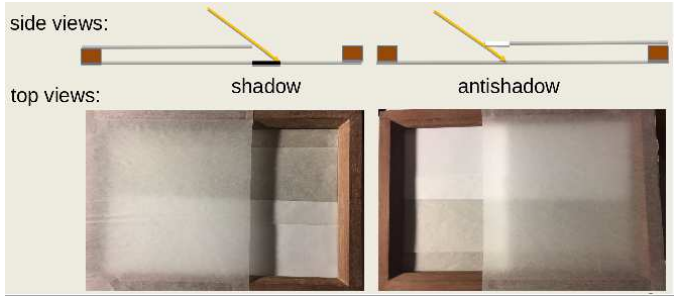}
\caption{A physical model illustrating the creation of shadows and
  antishadows by a translucent layer overlying a deeper scattering
  layer.  Here a wooden frame is used to suspend a sheet of tracing
  paper a short distance above several overlapping sheets that produce
  underlying albedo structure, which is visible through the overlying
  layer. Only part of the deeper layer is covered by the suspended
  sheet.  In both views light is incident at a low angle from the
  left.  On the left, the upper sheet covers about the left 60\% of the bottom
  sheets, while on the right it covers the right 60\%.  The result on
  the left is a shadow and, on the right, a bright region of similar
  width, which we call an antishadow.}\label{Fig:physmod2}
\end{figure}

It is also possible for optical depth changes in a translucent upper
aerosol layer to create shadows on an underlying layer, and to also
create bright features under appropriate illumination conditions.
This is illustrated by a physical model displayed in
Fig.\ \ref{Fig:physmod2}. The shadow is darker than both the covered
and uncovered regions away from the shadow because it is missing both
direct sunlight and the extra scattered light provided by the upper
layer.  In the bright region, in the right half of the panel, light is
able to provide extra illumination underneath the top layer which
increases its brightness above its normal value, seen further from the
edge, where there is no extra illumination.  Note that the bright
region, which we call an antishadow, has a width that is essentially
the same as that of the shadow in this case, and appears to be
a general feature of most of our trial calculations.

\begin{figure}[!hbt]\centering
\includegraphics[width=3.5in]{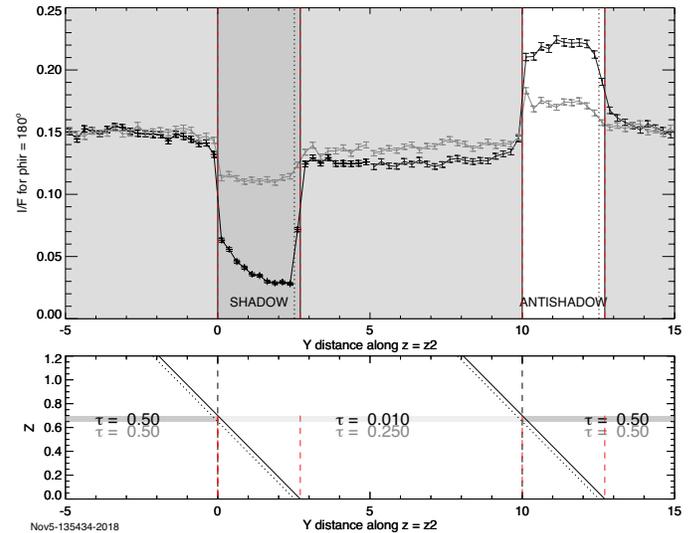}
\caption{Monte Carlo I/F calculation for a physically thin layer with
  step changes in opacity, above a Lambertian reflecting layer of
  albedo 0.5.  The incident light is coming from the left at $\mu_0$ =
  -0.25 at azimuth 90\deg (in the y-z plane).  The observer is at
  $\mu$= 0.55, with azimuth 180\degx (in the x-z plane).  The layer
  structure and observing and illumination geometry are illustrated in
  the bottom part of the figure, with the I/F profile in the upper
  portion. The particles in the layer are assumed to scatter conservatively with an
  HG phase function of asymmetry $g$ = 0.5 to simulate slightly
forward scattering particles. The shadow boundary is at
  $y = 0$, while the antishadow boundary is at $y=10$.  The distance
  scale is relative, but the same in $z$ and $y$ directions. Slanted
  solid and dotted lines in the bottom panel indicate incident rays,
  while dashed red lines indicate the projections of observer sight
  lines onto the y-z plane.}\label{Fig:mcthin}
\end{figure}

To determine how shadow depth and antishadow brightness depend on the
optical depth difference between two sides of an upper translucent
layer, we make use of the simplified Monte Carlo model described
previously. Fig.\ \ref{Fig:mcthin} displays the results of Monte Carlo
calculations for a physically thin layer above a Lambertian reflector.
The upper layer has a transition from $\tau_1$ to $\tau_2$ at y = 0
and back to $\tau_1$ at y=10.  We show two examples.  Both have
$\tau_1$ = 0.5. The first case, which is most similar to the physical
model, has the most dramatic transition, to $\tau_2$ = 0.01 and
back. In the second case the transition is to $\tau_2$ = 0.25 and
back.  The vertical structure, shown in the bottom panel, has the
upper scattering layer confined between z = 0.65 and z = 0.70, where
distance along y and z have the same arbitrary scale.  The more
dramatic transition displays a deep shadow that has its lowest I/F
displaced towards the outer limit of the shadow boundary as also seen
in the physical model.  The less dramatic transition produces flatter
and shallower shadows and antishadows.  The shapes of the shadow and
antishadow features also vary with viewing geometry and with the
thickness of the shadowing layer and its elevation above the lower
Lambertian reflecting layer.  Both features are also widened if the
transition itself occurs smoothly over a short distance rather than
as a step change.  However, detailed computations for gradual
transitions are much more complex to model and beyond the capabilities
of our current code.

\subsection{Quantitative comparisons with observed I/F profiles}

\begin{figure*}[!htb]\centering
\includegraphics[width=1.9in]{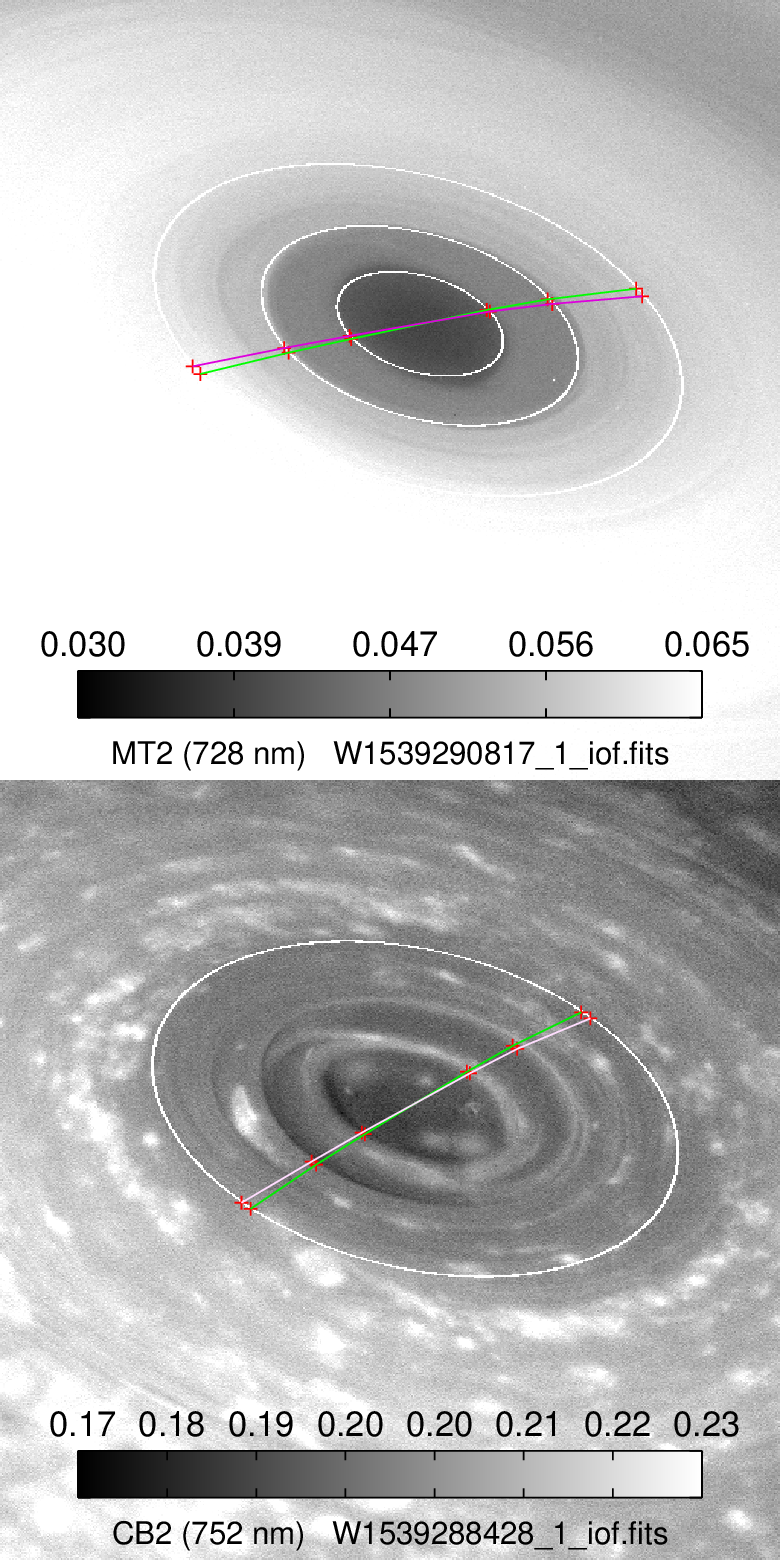}
\includegraphics[width=4.2in]{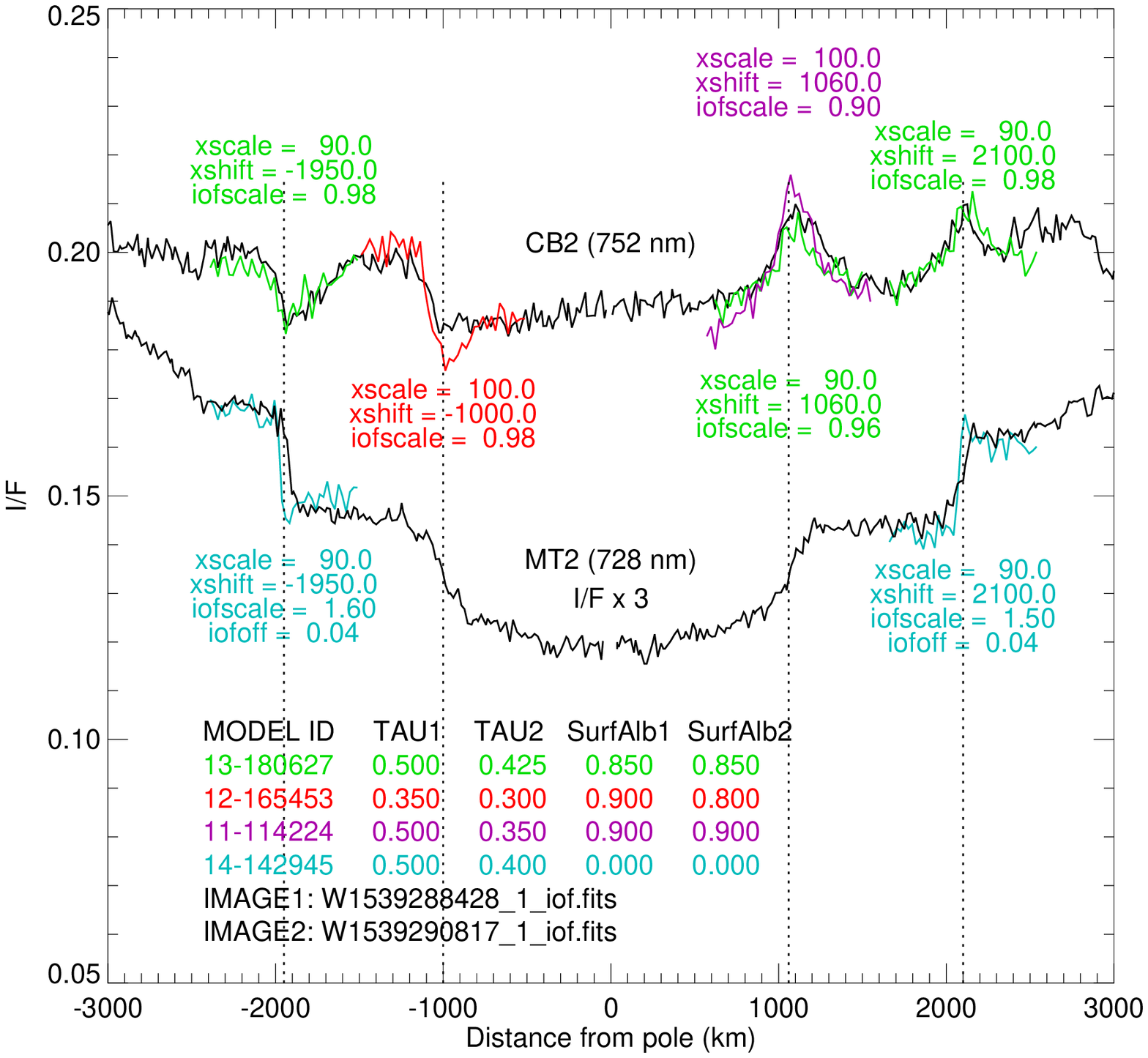}
\caption{Comparison of observed I/F profiles along lines indicated in
  images in the left two panels with Monte Carlo model profiles at and
  near boundaries where step changes in optical depth occur. The blue
  models are compared to the 728-nm I/F profile after multiplying the
  model length scale by the factor $xscale$ and boosting the I/F by
  the factor $yscale$. The $xshift$ parameter controls the horizontal
  position of the model I/F boundary and is adjusted to match the
  location of the observed I/F step. The I/F scale changes are
  equivalent to the same fractional adjustments needed in the size of
  the optical depth steps.}
\label{Fig:mcmodcomp}
\end{figure*}

Monte Carlo calculations for optical depth step changes of 0.05 to 0.1
from a base level of 0.35 to 0.5 were found to produce I/F amplitudes of about 10\%,
which is about the right amplitude to match the I/F variation seen
at 752 nm.  We also found that if we used similar transitions in the optical
depth of the diphosphine layer (near 250 mbar), we could match the I/F changes of 15\% to 20\%
seen at 728 nm, which is most sensitive to that layer. 
  Direct comparisons between I/F scans and Monte Carlo model
transitions are shown in Fig.\ \ref{Fig:mcmodcomp}. The absorption at 728 nm was
not included in the Monte Carlo calculation directly. Instead we assumed that
there was no contribution from below that layer due to that attenuation, then
just scaled the Monte Carlo I/F to account for absorption of incoming and outgoing
light by the methane above the layer.  

The shadow calculations shown in Fig.\ \ref{Fig:mcmodcomp} were
done in the relative scale domain.  The physical scale estimate was made
by finding a multiplication factor in km that produced approximately the correct
width of the calculated features.  With our cloud tops at 0.9 relative units
and scale factors of typically 90 to 100, this led to rather large distances
between the shadowed layer and the shadow-casting layer, of typically 80-90
km.  That is more altitude separation than the 60-70 km that exists between
the stratosphere and diphosphine layers, but less than the altitude
separation between the stratosphere and the \nht layer.  Given the
semi-transparent nature of the \pthf layer, is is conceivable that the
observed shadow is actually a combination of shadows on both layers.
Unfortunately, our Monte Carlo code is too simple to handle the shadowing
of one layer onto two lower layers.  

More problematic is the situation for the inner transition near 1.1\deg from the
pole.  In this case the model needs a similar 80-90 km vertical separation between
layers, but only about half that distance exists between the diphosphine
layer with the transition and the underlying ammonia layer (see Fig.\ \ref{Fig:vimsfits}).  However, the ammonia
layer itself may be sufficiently transparent that a shadow falling through
the ammonia layer onto the deep layer might contribute to the observed
length of the inner shadow.  Even though its optical depth is about 0.6
at 752 nm, its larger particles will forward scatter, so that they will
be more translucent that might be initially expected.
  The deeper layer is of the order of 100 km below
the diphosphine layer.

\section{Evaluation of shadow and antishadow possibilities}

\begin{figure*}[!htb]\centering
\includegraphics[width=2.1in]{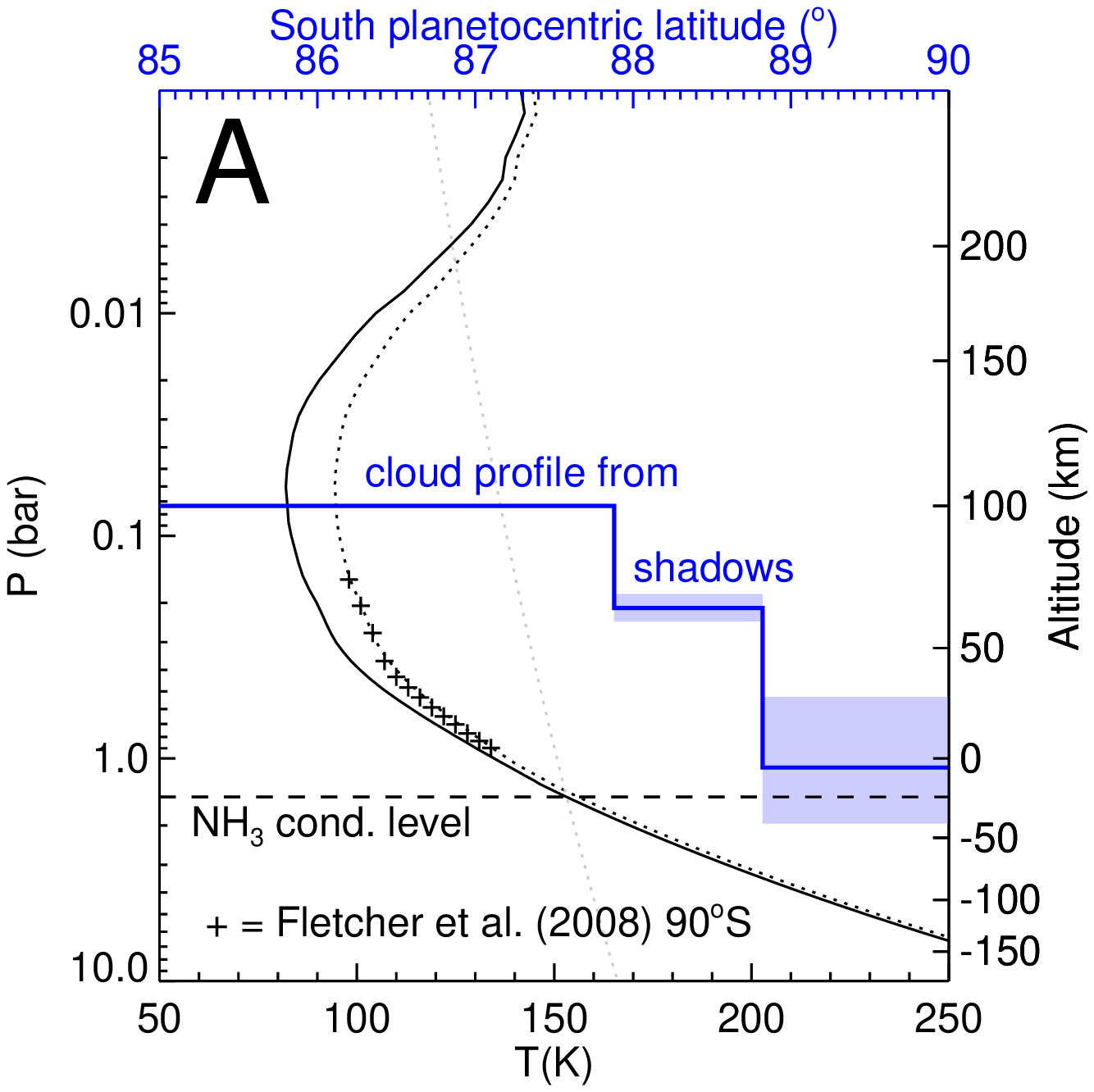}\hspace{-0.1in}
\includegraphics[width=2.1in]{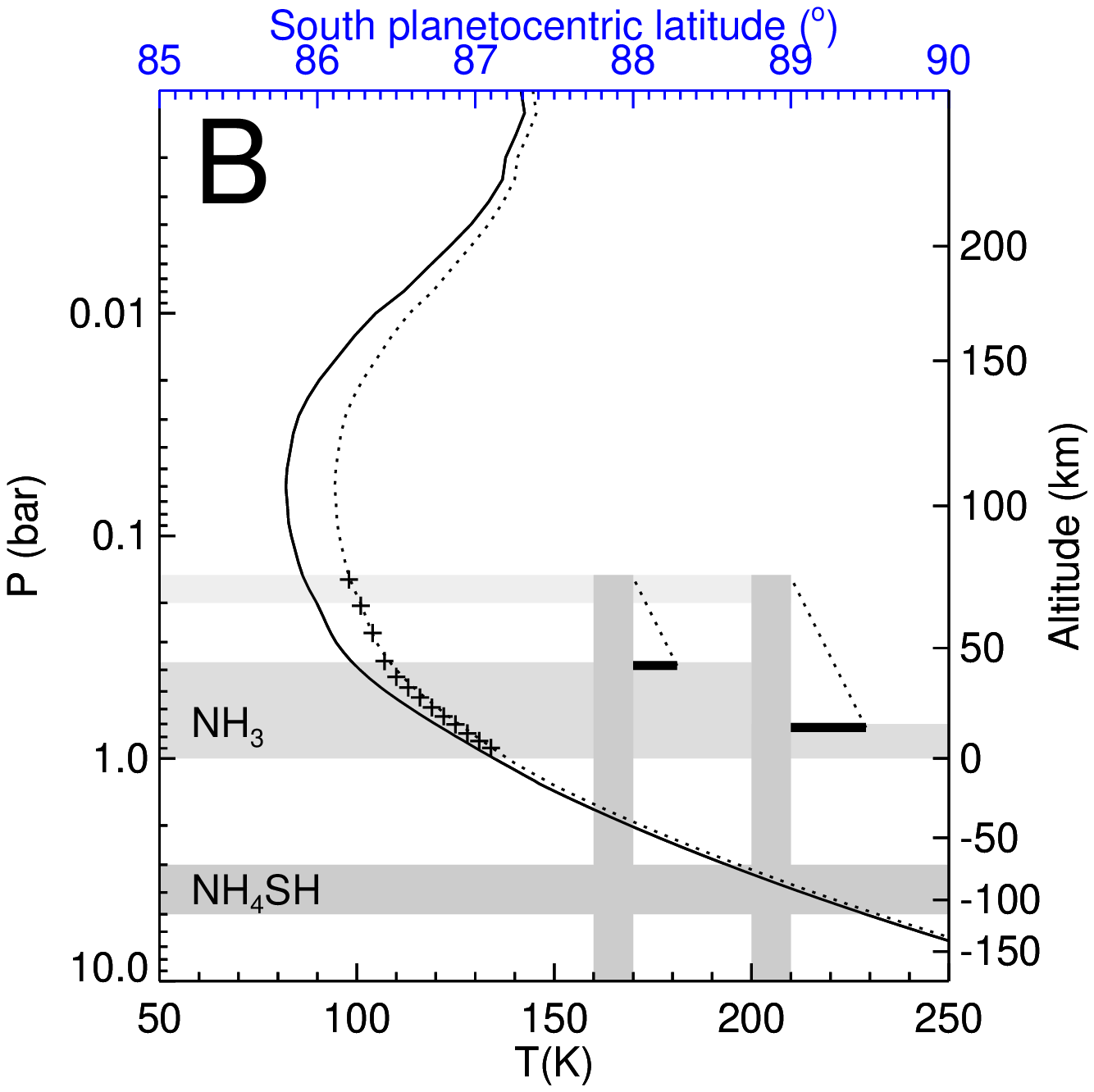}\hspace{-0.1in}
\includegraphics[width=2.1in]{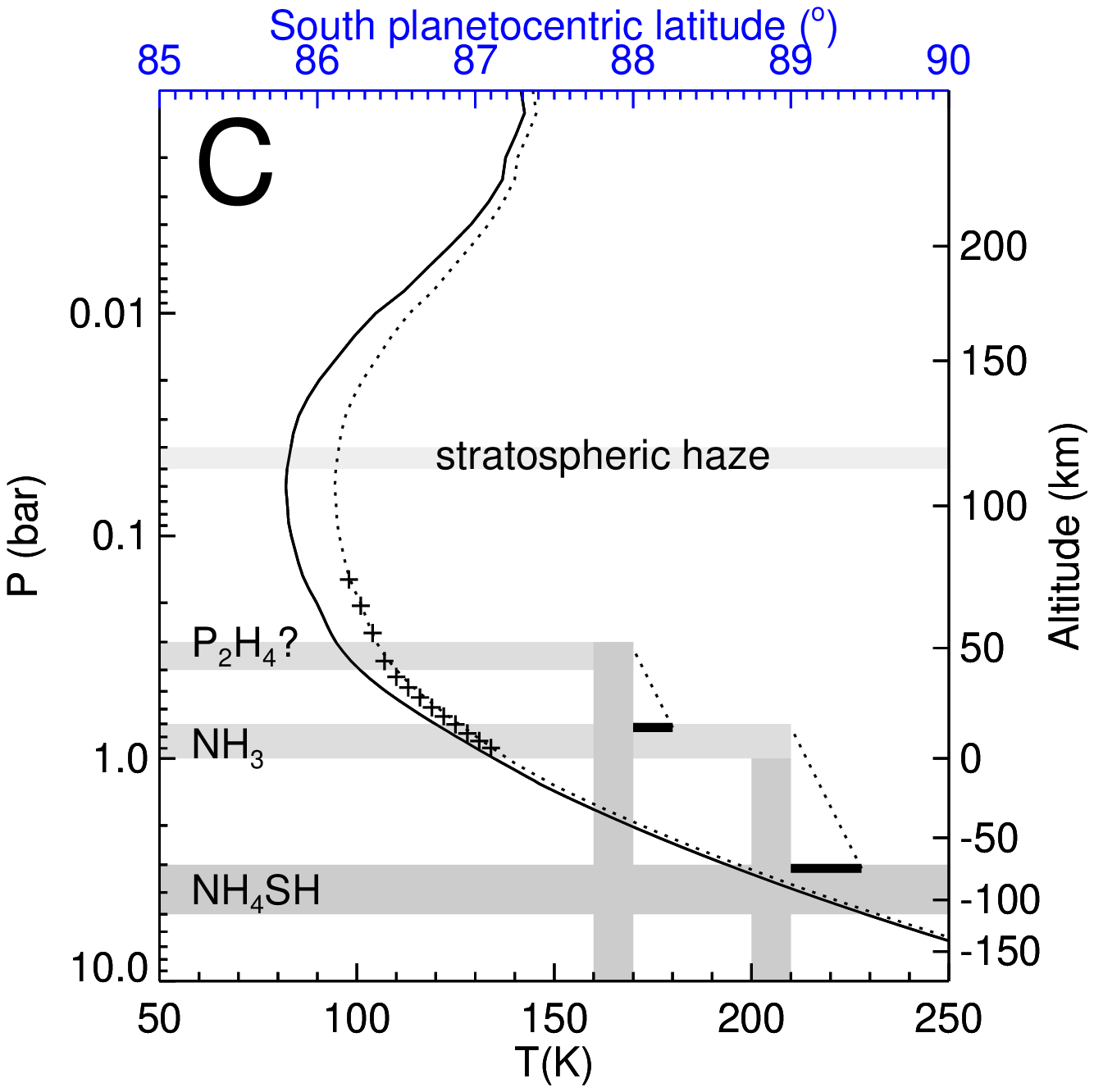}
\includegraphics[width=2.1in]{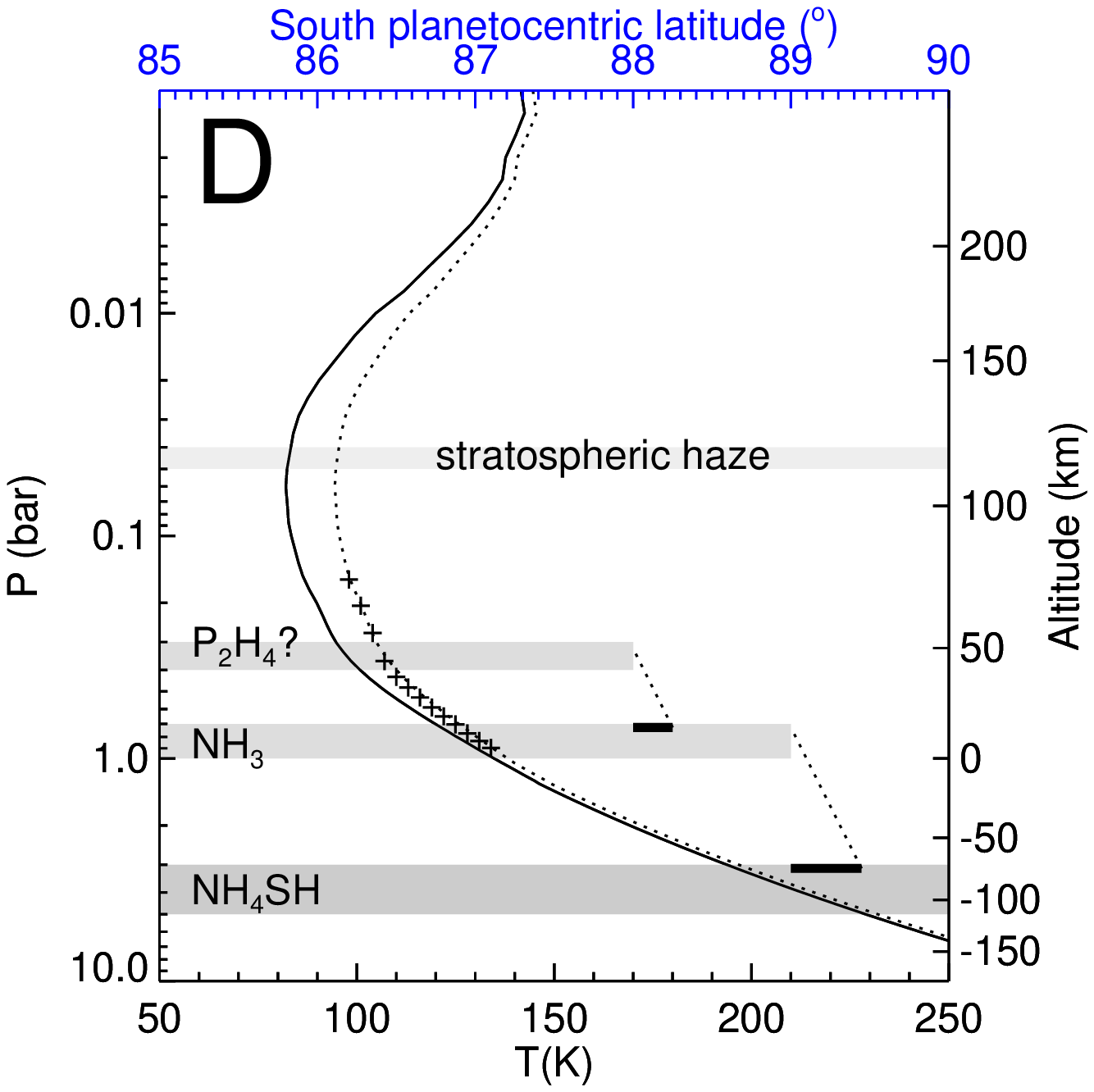}\hspace{-0.1in}
\includegraphics[width=2.1in]{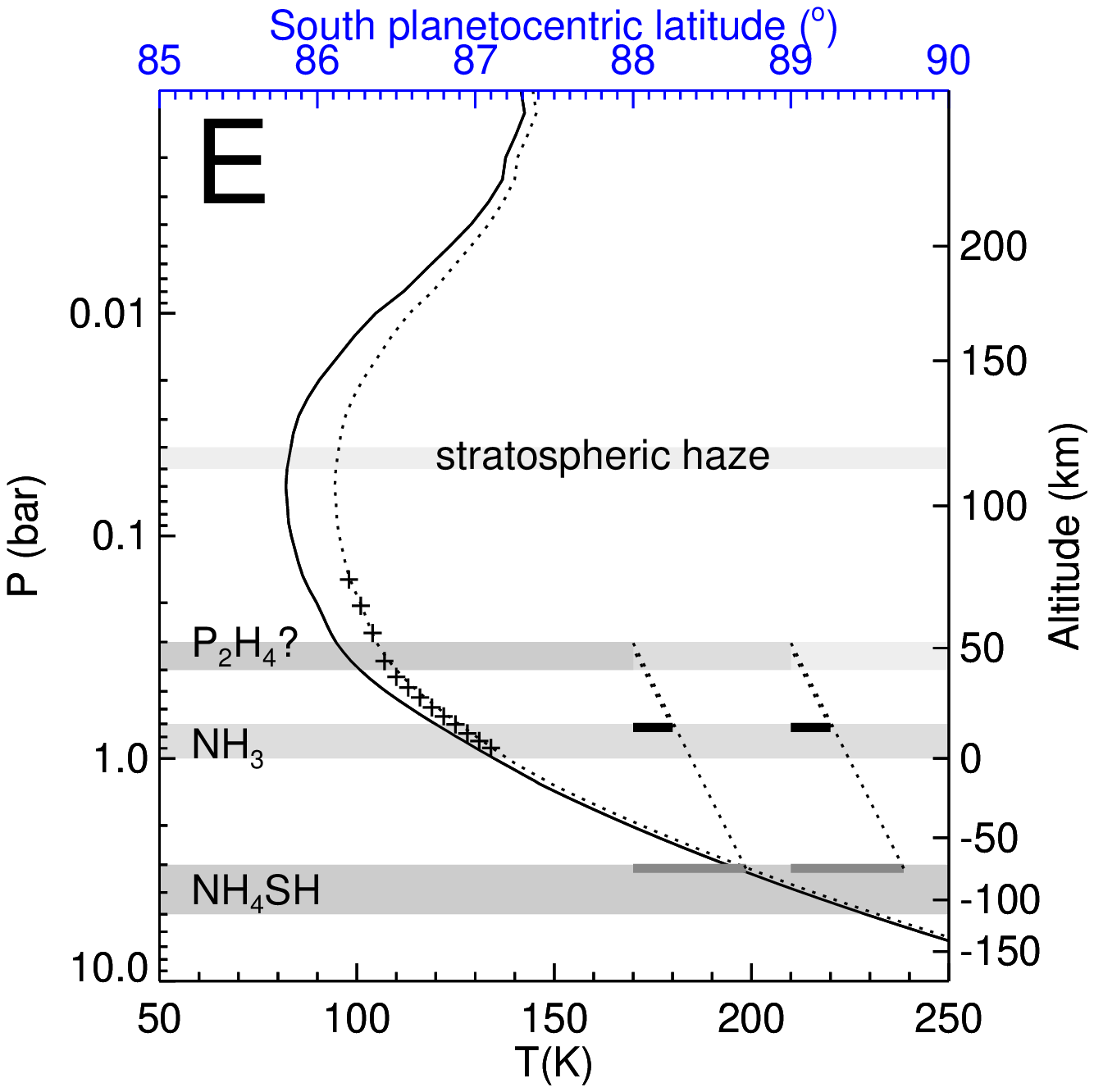}\hspace{-0.1in}
\includegraphics[width=2.1in]{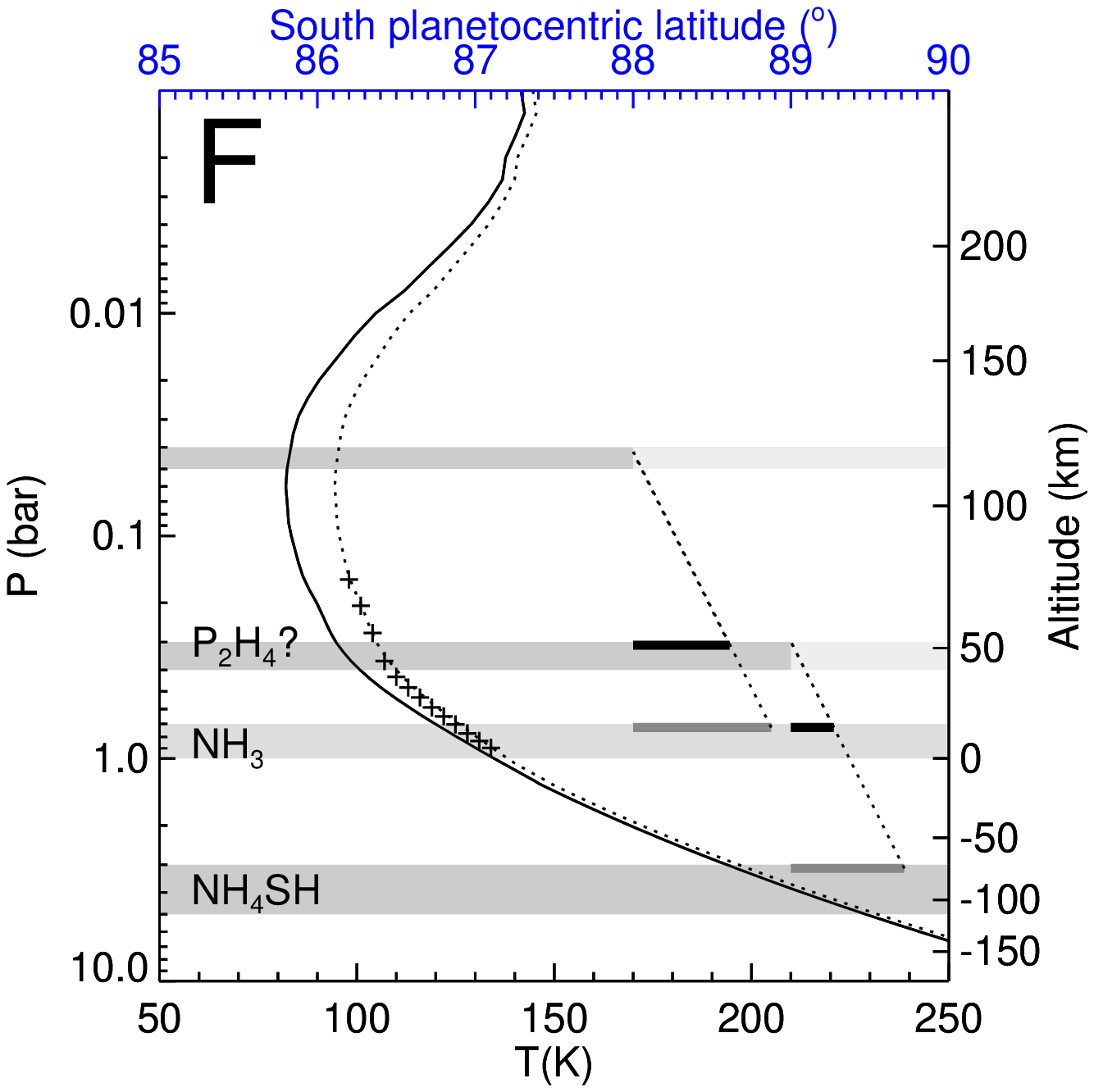}
\caption{Six candidate cloud profiles that might be at least crudely  consistent with shadow measurements by \cite{Dyudina2009}. Panel A
  starts with an outer cloud at the tropopause level, which must be 30
  km above the cloud level on which it casts shadows,
 and that layer is assumed to cast shadows on a layer that
  is roughly 70 km deeper, which would put it near the 1 bar level.  Examples
  B and C contain deeply convective ``eyewall'' clouds that cast
  shadows on the \nht layer, or first on the \nht layer and then on
  the \nhfsh layer.  In panel D, the shadows are cast by upper layers
  on lower layers without deeply convective wall clouds.  In panel E,
  the step changes in the optical depth of the putative diphosphine
  layer casts both inner and outer shadows on the ammonia layer, and
  in panel F, an optical depth step in the stratospheric haze casts
a shadow on diphosphine and ammonia layers and a step change in the
diphosphine layer casts a shadow on the ammonia and NH$_4$SH layers.}\label{Fig:profileoptions}
\end{figure*}

Six different shadow-generating vertical cloud profiles are plotted in
Fig.\ \ref{Fig:profileoptions}.  We consider each of these cases
and weigh the evidence that is relevant.

\vspace{0.05in} 
\noindent
{\bf Panel A (stair steps of optically thick clouds). } This starts with clouds
tops reaching above the tropopause, declining about 30 km near 88\degx S, then
declining another 70$\pm$30 km near 89\degx S, with no overlying cloud layers.
This can be rejected because VIMS spectral constraints do not find optically
thick clouds or any significant pressure changes in any of the four cloud layers,
and especially no layers in which clouds are completely cleared out.
Also, there are no observations of bright eyewalls that should be produced
by such a structure, and no mechanism to create antishadows that are observed.
Furthermore, the shadows from such structures would be very dark, not
just 10\% perturbations of the I/F profile with latitude.

\vspace{0.05in} 
\noindent
{\bf Panel B (two deep convective eyewall clouds casting shadows on
a background ammonia layer). }  In this case different shadow lengths
arise from different heights of inner and outer wall clouds.   A thick
cirrus outflow away from the pole must be rejected because it would
obscure the shadow from the outer eyewall.  But for a thin outflow
layer, shadows should be seen on both sides of the pole, unless the
eyewalls have a shallow slope extending a long distance away from the
pole.  But VIMS constraints indicate that all layers slope downward toward
the pole instead of away from it.  Also, there is no evidence in the
spectra for deeply convective optically thick clouds, as well as
no observations of lightning that might be expected from such convection,
as seen elsewhere on Saturn.  Further, this structure provides no
mechanism for producing antishadows.

\vspace{0.05in} 
\noindent
{\bf Panel C (two deep convective eyewalls casting shadows on
  different layers). }  In this case the inner eyewall casts a shadow
on the putative \nhfsh layer and generates an ammonia ice outflow
layer on which the outer eyewall casts its shadows.  If the outflow
layers are thick, this would prevent forming the second set of shadows
on the side of the pole opposite where poleward shadows are created.
But these should produce bright eyewalls, which are not seen, and
cannot produce antishadows, which are seen. Also, VIMS spectra provide
no evidence for optically thick clouds in this region, and also no
clear regions interior to the shadow forming cloud boundaries.

\vspace{0.15in} 
\noindent
{\bf Panel D (\pthf layer casting outer shadow on \nht layer which
  casts inner shadow on \nhfsh layer).} In this case two cloud layers
simply disappear in succession as a function of increasing latitude
(toward the pole).  The top layer would then cast a shadow on the next
layer down, and finally the second layer would cast a shadow on the
third layer. This has the virtue of providing height differentials that
are consistent with shadow length measurements of \cite{Dyudina2009},
and is also capable of producing both shadows and antishadows.  But
the successive clearing of upper layers is inconsistent with VIMS
spectral results, and the large optical depth changes by clearing
transitions would produce much deeper shadows and brighter antishadows
than are observed.

\vspace{0.05in} 
\noindent
{\bf Panel E (the \pthf layer produces both inner and outer shadows on the
\nht layer by step changes in optical depth). } This was the model that
seemed most consistent with our initial VIMS analysis \citep{Sro2018DPS} for
which the stratospheric haze was forced to be at too low a pressure.
Our current analysis indicates that the stratospheric haze has a more
significant optical depth and is likely
the location of the outer step in optical depth, with the putative
\pthf layer being the most probable location of the inner optical depth
transition. This structure also is problematic in having too little distance between
layers that create the inner shadow, requiring that the shadow length be extended perhaps by an attenuated
shadow on the deeper layer.  A quantitative evaluation of this
 of this multi-layer
shadow is beyond the capabilities of our current Monte Carlo code.  

\vspace{0.05in} 
\noindent
{\bf Panel F (a step change in the stratospheric haze optical depth casts a shadow primarily on the \pthf layer, 
and a similar optical depth step in the putative \pthf casts a shadow primarily on the \nht layer, but also
on the deeper putative \nhfsh layer). }  This has the best consistency with VIMS spectra,
and at least the outer transition is supported by our Monte Carlo calculations, but the inner model shadow
on the \nht layer is too short to fit the observations.  The longer shadow that might fall on the deeper
layer might produce a better match to the observations, but our Monte Carlo model is insufficiently complex to treat the
combination of all three layers, and reach a definitive conclusion about the inner shadow.

\vspace{0.1in} 
\noindent
The weight of the evidence favors the production of shadows by
relatively sharp optical depth transitions with latitude, the outermost being in the
stratospheric haze layer and the innermost being in the putative diphosphine layer.
This mechanism
can produce both shadows and antishadows of approximately the correct
amplitude, and does not require deep convective clouds that are not
observed, and thus the lack of lightning detections is also consistent
with this structure.


\section{Summary and Conclusions}

Our analysis of Cassini/VIMS spectra of the south polar region of Saturn have led to the following conclusions.

\begin{enumerate}

\item The cloud structures inferred from radiative transfer modeling
  are not consistent with the step change in cloud altitudes derived
  from cloud shadows by \cite{Dyudina2009}.  None of the modeled cloud
  layers disappear, and their optical depths are relatively small and
  do not change dramatically with latitude. There is no evidence for
  an optically thick vertical wall of clouds as found in hurricanes on
  Earth.

\item Monte Carlo calculations confirm that illuminated eyewall clouds
  should be very bright and that these bright features should appear
  to extend poleward from the wall boundary as would
  the shadows that such a wall would produce with the sun in the
opposite direction.  But this contradicts
  the finding of slightly bright features extending
  away from the pole (Fig.\ \ref{Fig:closeup}), which is the
  expected direction for antishadows.

\item The shadow boundaries do correlate with changes in scattering
  properties of overlying hazes, which appear to involve
  two step changes in optical depth, one in the stratospheric haze
  near 2.1\deg from the south pole, and one in the putative
diphosphine haze near 1.1\deg from the pole. 
 According to Monte Carlo calculations, these steps are of sufficient
magnitude to produce weak shadows, and weak shadows are observed
  (Fig.\ \ref{Fig:mcmodcomp}).

\item The shadows produced by the overlying hazes have the
  appropriate physical configuration to explain the anti-shadows
  observed on the opposite side of the pole from the shadows.  These are
  regions of local brightness increases produced by sunlight
  illuminating the layer underneath the shadow casting layer,
  causing it to appear brighter from above than the region further
  from the pole that does not receive that extra source of light.

\end{enumerate}

\noindent
There remain a number of questions to be resolved about these
features.  A better constraint on the cloud structure might be
obtained from more complex Monte Carlo calculations that can handle
more complex and more realistic vertical variations, including
multi-layer shadows and anti-shadows.  It would also be interesting to
apply this analysis to the north polar region.  It is curious that in
spite of large decreases in optical depths of haze and cloud layers as
the north pole is approached \citep{Baines2018GeoRL}, prominent cloud
shadows have not been noted at high north polar latitudes, perhaps
because the layers on which those shadows might be cast are also of
low optical depth or because the transitions are insufficiently
sharp. That situation could benefit from further
investigation. There also remain significant uncertainties regarding
the composition of the ubiquitous upper tropospheric layer, which
seems likely to be either diphosphine or some form of phosphorus. But
to test which better fits the spectral observations, we need better
measurements of the optical properties of both of these materials
under Saturnian conditions, especially of diphosphine about which very
little is known.

\section*{Acknowledgments.} \addcontentsline{toc}{section}{Acknowledgments}

Support for this work was provided by
NASA through its Cassini Data Analysis and
Participating Scientists Program via grant NNX15AL10G and by
the subsequent Cassini Data Analysis Program via grant 80NSSC18K0966.
We thank Robert West for a prompt and constructive review.
The archived data associated with the paper can be obtained from the 
Planetary Data System (PDS) Atmospheres Node,
and includes calibrated Cassini VIMS and ISS datasets, along with a 
description of the calibration, and tabular data for all tables and 
figures contained in the paper as well as a public domain 
version of the paper. 


\begin{thebibliography}{18}
\expandafter\ifx\csname natexlab\endcsname\relax\def\natexlab#1{#1}\fi
\expandafter\ifx\csname url\endcsname\relax
  \def\url#1{\texttt{#1}}\fi
\expandafter\ifx\csname urlprefix\endcsname\relax\def\urlprefix{URL }\fi

\bibitem[{{Antu{\~n}ano} et~al.(2015){Antu{\~n}ano},
  {R{\'{\i}}o-Gaztelurrutia}, {S{\'a}nchez-Lavega}, and {Hueso}}]{Antunano2015}
{Antu{\~n}ano}, A., {R{\'{\i}}o-Gaztelurrutia}, T., {S{\'a}nchez-Lavega}, A.,
  {Hueso}, R., 2015. {Dynamics of Saturn's polar regions}. J. of Geophys. Res.
  (Planets) 120, 155--176.

\bibitem[{{Baines} et~al.(2009{\natexlab{a}}){Baines}, {Delitsky}, {Momary},
  {Brown}, {Buratti}, {Clark}, and {Nicholson}}]{Baines2009stormclouds}
{Baines}, K.~H., {Delitsky}, M.~L., {Momary}, T.~W., {Brown}, R.~H., {Buratti},
  B.~J., {Clark}, R.~N., {Nicholson}, P.~D., 2009{\natexlab{a}}. {Storm clouds
  on Saturn: Lightning-induced chemistry and associated materials consistent
  with Cassini/VIMS spectra}. \planss 57, 1650--1658.

\bibitem[{{Baines} et~al.(2009{\natexlab{b}}){Baines}, {Momary}, {Fletcher},
  {Showman}, {Roos-Serote}, {Brown}, {Buratti}, {Clark}, and
  {Nicholson}}]{Baines2009cyclone}
{Baines}, K.~H., {Momary}, T.~W., {Fletcher}, L.~N., {Showman}, A.~P.,
  {Roos-Serote}, M., {Brown}, R.~H., {Buratti}, B.~J., {Clark}, R.~N.,
  {Nicholson}, P.~D., 2009{\natexlab{b}}. {Saturn's north polar cyclone and
  hexagon at depth revealed by Cassini/VIMS}. \planss 57, 1671--1681.

\bibitem[{{Baines} et~al.(2018){Baines}, {Sromovsky}, {Fry}, {Momary}, {Brown},
  {Buratti}, {Clark}, {Nicholson}, and {Sotin}}]{Baines2018GeoRL}
{Baines}, K.~H., {Sromovsky}, L.~A., {Fry}, P.~M., {Momary}, T.~W., {Brown},
  R.~H., {Buratti}, B.~J., {Clark}, R.~N., {Nicholson}, P.~D., {Sotin}, C.,
  2018. {The Eye of Saturn's North Polar Vortex: Unexpected Cloud Structures
  Observed at High Spatial Resolution by Cassini/VIMS}. \grl 45, 5867--5875.

\bibitem[{{Brown} et~al.(2004){Brown}, {Baines}, {Bellucci}, {Bibring},
  {Buratti}, {Capaccioni}, {Cerroni}, {Clark}, {Coradini}, {Cruikshank},
  {Drossart}, {Formisano}, {Jaumann}, {Langevin}, {Matson}, {McCord},
  {Mennella}, {Miller}, {Nelson}, {Nicholson}, {Sicardy}, and
  {Sotin}}]{Brown2004}
{Brown}, R.~H., {Baines}, K.~H., {Bellucci}, G., {Bibring}, J.-P., {Buratti},
  B.~J., {Capaccioni}, F., {Cerroni}, P., {Clark}, R.~N., {Coradini}, A.,
  {Cruikshank}, D.~P., {Drossart}, P., {Formisano}, V., {Jaumann}, R.,
  {Langevin}, Y., {Matson}, D.~L., {McCord}, T.~B., {Mennella}, V., {Miller},
  E., {Nelson}, R.~M., {Nicholson}, P.~D., {Sicardy}, B., {Sotin}, C., 2004.
  {The Cassini Visual and Infrared Mapping Spectrometer (VIMS) Investigation}.
  Space Sci. Rev. 115, 111--168.

\bibitem[{{Dyudina} et~al.(2007){Dyudina}, {Ingersoll}, {Ewald}, {Porco},
  {Fischer}, {Kurth}, {Desch}, {Del Genio}, {Barbara}, and
  {Ferrier}}]{Dyudina2007}
{Dyudina}, U.~A., {Ingersoll}, A.~P., {Ewald}, S.~P., {Porco}, C.~C.,
  {Fischer}, G., {Kurth}, W., {Desch}, M., {Del Genio}, A., {Barbara}, J.,
  {Ferrier}, J., 2007. {Lightning storms on Saturn observed by Cassini ISS and
  RPWS during 2004--2006}. Icarus 190, 545--555.

\bibitem[{{Dyudina} et~al.(2013){Dyudina}, {Ingersoll}, {Ewald}, {Porco},
  {Fischer}, and {Yair}}]{Dyudina2013}
{Dyudina}, U.~A., {Ingersoll}, A.~P., {Ewald}, S.~P., {Porco}, C.~C.,
  {Fischer}, G., {Yair}, Y., 2013. {Saturn's visible lightning, its radio
  emissions, and the structure of the 2009-2011 lightning storms}. Icarus 226,
  1020--1037.

\bibitem[{{Dyudina} et~al.(2009){Dyudina}, {Ingersoll}, {Ewald}, {Vasavada},
  {West}, {Baines}, {Momary}, {Del Genio}, {Barbara}, {Porco}, {Achterberg},
  {Flasar}, {Simon-Miller}, and {Fletcher}}]{Dyudina2009}
{Dyudina}, U.~A., {Ingersoll}, A.~P., {Ewald}, S.~P., {Vasavada}, A.~R.,
  {West}, R.~A., {Baines}, K.~H., {Momary}, T.~W., {Del Genio}, A.~D.,
  {Barbara}, J.~M., {Porco}, C.~C., {Achterberg}, R.~K., {Flasar}, F.~M.,
  {Simon-Miller}, A.~A., {Fletcher}, L.~N., 2009. {Saturn's south polar vortex
  compared to other large vortices in the Solar System}. Icarus 202, 240--248.

\bibitem[{{Dyudina} et~al.(2008){Dyudina}, {Ingersoll}, {Ewald}, {Vasavada},
  {West}, {Del Genio}, {Barbara}, {Porco}, {Achterberg}, {Flasar},
  {Simon-Miller}, and {Fletcher}}]{Dyudina2008Sci}
{Dyudina}, U.~A., {Ingersoll}, A.~P., {Ewald}, S.~P., {Vasavada}, A.~R.,
  {West}, R.~A., {Del Genio}, A.~D., {Barbara}, J.~M., {Porco}, C.~C.,
  {Achterberg}, R.~K., {Flasar}, F.~M., {Simon-Miller}, A.~A., {Fletcher},
  L.~N., 2008. {Dynamics of Saturn's South Polar Vortex}. Science 319, 1801.

\bibitem[{{Fletcher} et~al.(2008){Fletcher}, {Irwin}, {Orton}, {Teanby},
  {Achterberg}, {Bjoraker}, {Read}, {Simon-Miller}, {Howett}, {de Kok},
  {Bowles}, {Calcutt}, {Hesman}, and {Flasar}}]{Fletcher2008}
{Fletcher}, L.~N., {Irwin}, P.~G.~J., {Orton}, G.~S., {Teanby}, N.~A.,
  {Achterberg}, R.~K., {Bjoraker}, G.~L., {Read}, P.~L., {Simon-Miller}, A.~A.,
  {Howett}, C., {de Kok}, R., {Bowles}, N., {Calcutt}, S.~B., {Hesman}, B.,
  {Flasar}, F.~M., 2008. {Temperature and Composition of Saturn's Polar Hot
  Spots and Hexagon}. Science 319, 79.

\bibitem[{{S{\'a}nchez-Lavega} et~al.(2006){S{\'a}nchez-Lavega}, {Hueso},
  {P{\'e}rez-Hoyos}, and {Rojas}}]{Sanchez-Lavega2006}
{S{\'a}nchez-Lavega}, A., {Hueso}, R., {P{\'e}rez-Hoyos}, S., {Rojas}, J.~F.,
  2006. {A strong vortex in Saturn's South Pole}. Icarus 184, 524--531.

\bibitem[{{Sromovsky} et~al.(2013){Sromovsky}, {Baines}, and
  {Fry}}]{Sro2013gws}
{Sromovsky}, L.~A., {Baines}, K.~H., {Fry}, P.~M., 2013. {Saturn's Great Storm
  of 2010-2011: Evidence for ammonia and water ices from analysis of VIMS
  spectra}. Icarus 226, 402--418.

\bibitem[{{Sromovsky} et~al.(2018){Sromovsky}, {Baines}, and
  {Fry}}]{Sro2018dark}
{Sromovsky}, L.~A., {Baines}, K.~H., {Fry}, P.~M., 2018. {Models of bright
  storm clouds and related dark ovals in Saturn's Storm Alley as constrained by
  2008 Cassini/VIMS spectra}. Icarus 302, 360--385.

\bibitem[{{Sromovsky} et~al.(2019){Sromovsky}, {Baines}, and
  {Fry}}]{Sro2019spole}
{Sromovsky}, L.~A., {Baines}, K.~H., {Fry}, P.~M., 2019. {Saturn's south polar
  cloud structure inferred from 2006 Cassini VIMS spectra}. Icarus, submitted.

\bibitem[{{Sromovsky} et~al.(2016){Sromovsky}, {Baines}, {Fry}, and
  {Momary}}]{Sro2016}
{Sromovsky}, L.~A., {Baines}, K.~H., {Fry}, P.~M., {Momary}, T.~W., 2016.
  {Cloud clearing in the wake of Saturn's Great Storm of 2010-2011 and
  suggested new constraints on Saturn's He/H$_{2}$ ratio}. Icarus 276,
  141--162.

\bibitem[{{Sromovsky} and {Fry}(2018)}]{Sro2018DPS}
{Sromovsky}, L.~A., {Fry}, P.~M., 2018. {The Nature of South Polar Cloud
  Shadows and Anti-Shadows on Saturn}. In: AAS/Division for Planetary Sciences
  Meeting Abstracts. Vol.~50 of AAS/Division for Planetary Sciences Meeting
  Abstracts. p. 507.05.

\bibitem[{{Whitney}(2011)}]{Whitney2011}
{Whitney}, B.~A., 2011. {Monte Carlo radiative transfer}. Bulletin of the
  Astronomical Society of India 39, 101--127.

\bibitem[{{Witt}(1977)}]{Witt1977}
{Witt}, A.~N., 1977. {Multiple scattering in reflection nebulae. I - A Monte
  Carlo approach.} \apjs 35, 1--6.

\end{thebibliography}

\end{document}